\documentclass[a4paper,11pt]{article}
\pdfoutput=1 
\usepackage{jheppub_forArxiv}

\hypersetup{pdfauthor={Jannik Hofest{\"a}dt},pdftitle={Intrinsic limits on resolutions in muon- and electron-neutrino charged-current events in the KM3NeT/ORCA detector}}

\usepackage{aas_macros}
\usepackage[T1]{fontenc}

\usepackage{lineno}

\usepackage{amsmath}
\usepackage{amssymb}

\usepackage{graphicx}

\usepackage{color}
\usepackage{vmargin}
\usepackage[utf8]{inputenc}
\usepackage[T1]{fontenc}
\usepackage{ulem}

\usepackage{hyperref}

\catcode`\@=11
\def\parenbar{\mathpalette\p@renb@r}
\def\p@renb@r#1#2{\vbox{%
  \ifx#1\scriptscriptstyle \dimen@.7em\dimen@ii.2em\else
  \ifx#1\scriptstyle \dimen@.8em\dimen@ii.25em\else
  \dimen@1em\dimen@ii.4em\fi\fi \offinterlineskip
  \ialign{\hfill##\hfill\cr
    \vbox{\hrule width\dimen@ii}\cr
    \noalign{\vskip-.3ex}%
    \hbox to\dimen@{$\mathchar300\hfil\mathchar301$}\cr
    \noalign{\vskip-.3ex}%
    $#1#2$\cr}}}
\catcode`\@=12
\def\nuan{\parenbar{\nu}\kern-0.4ex}

\title{Intrinsic limits on resolutions in muon- and electron-neutrino charged-current events in the KM3NeT/ORCA detector}

\collaborationImg{\includegraphics{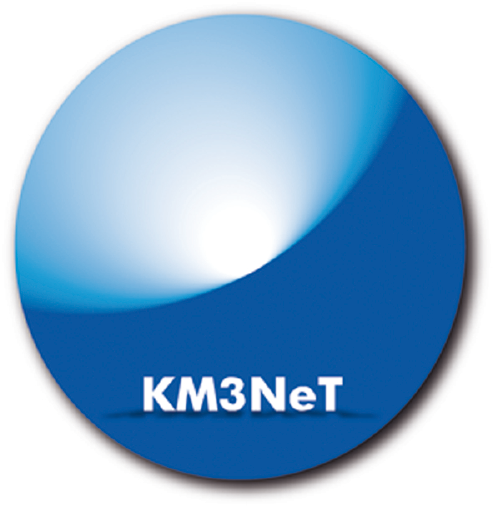}}

\author[ao]{S.~Adri{\'a}n-Mart{\'\i}nez,}
\author[b]{M.~Ageron,}
\author[s]{S.~Aiello,}
\author[ap]{A.~Albert,}
\author[w]{F.~Ameli,}
\author[ad]{E.~G.~Anassontzis,}
\author[an]{M.~Andre,}
\author[ab]{G.~Androulakis,}
\author[t]{M.~Anghinolfi,}
\author[g]{G.~Anton,}
% ...
\affiliation[b]{Aix-Marseille Universit{\'e},~CNRS/IN2P3,~CPPM~UMR~7346,~13288,~Marseille,~France}
% ...
\affiliation[g]{Friedrich-Alexander-Universit{\"a}t Erlangen-N{\"u}rnberg, Erlangen Centre for Astroparticle Physics, Erwin-Rommel-Stra{\ss}e 1, 91058 Erlangen, Germany}
% ...
\affiliation[s]{INFN, Sezione di Catania, Via Santa Sofia 64, Catania, 95123 Italy}
\affiliation[t]{INFN, Sezione di Genova, Via Dodecaneso 33, Genova, 16146 Italy}
% ...
\affiliation[w]{INFN, Sezione di Roma, Piazzale Aldo Moro 2, Roma, 00185 Italy}
% ...
\affiliation[ab]{NCSR Demokritos, Institute of Nuclear and Particle Physics, Ag. Paraskevi Attikis, Athens, 15310 Greece}
% ...
\affiliation[ad]{Physics~Department,~N.~and~K.~University~of~Athens,~Athens,~Greece}
% ...
\affiliation[an]{Universitat Polit{\`e}cnica de Catalunya, Laboratori d'Aplicacions Bioac{\'u}stiques, Centre Tecnol{\`o}gic de Vilanova i la Geltr{\'u}, Avda. Rambla Exposici{\'o}, s/n, Vilanova i la Geltr{\'u}, 08800 Spain}
\affiliation[ao]{Universitat Polit{\`e}cnica de Val{\`e}ncia, Instituto de Investigaci{\'o}n para la Gesti{\'o}n Integrada de las Zonas Costeras, C/ Paranimf, 1, Gandia, 46730 Spain}
\affiliation[ap]{Universit{\'e} de Strasbourg, Universit{\'e} de Haute Alsace, GRPHE, 34, Rue du Grillenbreit, Colmar, 68008 France}
% ...

\emailAdd{jannik.hofestaedt@fau.de}
\emailAdd{clancy.james@fau.de}

\abstract{Studying atmospheric neutrino oscillations in the few-GeV range with a multi-megaton detector promises to determine the neutrino mass hierarchy. This is the main science goal pursued by the future KM3NeT/ORCA water Cherenkov detector in the Mediterranean Sea. In this paper, the processes that limit the obtainable resolution in both energy and direction in charged-current neutrino events in the ORCA detector are investigated. These processes include the composition of the hadronic fragmentation products, the subsequent particle propagation and the photon-sampling fraction of the detector. GEANT simulations of neutrino interactions in seawater produced by GENIE are used to study the effects in the 1--20\,GeV range.
It is found that fluctuations in the hadronic cascade in conjunction with the variation of the inelasticity $y$ are most detrimental to the resolutions. The effect of limited photon sampling in the detector is of significantly less importance. These results will therefore also be applicable to similar detectors/media, such as those in ice.}

\arxivnumber{1612.05621}

\begin{document}

\maketitle
\flushbottom

\section{Introduction}
\label{sec:introduction}

It has recently been suggested that the neutrino mass hierarchy can be determined with multi-megaton neutrino detectors in water and ice by measuring the flux of atmospheric neutrinos that have passed through the Earth~\cite{2013JHEP...02..082A}. Due to the influence of matter on neutrino oscillation during propagation through the Earth, the expected interaction rate of neutrinos in the energy regime of $\sim1$--$20$\,GeV differs between the normal and inverted hierarchies. This is the main science goal pursued by KM3NeT/ORCA, the densely instrumented component of KM3NeT 2.0 being built off the southern coast of France~\cite{KM3NeT_LoI}. In its current design, ORCA will consist of a 3D array of $2070$ optical modules, each housing $31$ $3$-inch photomultiplier tubes (PMTs), with about $20$\,m horizontal and $9$\,m vertical spacing between optical modules, instrumenting in total a volume of about $5.5 \times 10^6\,\text{m}^3$ of seawater. The PINGU infill array of the IceCube detector at the South Pole has also been proposed to perform this measurement~\cite{PINGU}. Additionally, such detectors are expected to further constrain neutrino-oscillation parameters, and have been proposed to study sterile neutrinos, non-standard interactions, dark matter, supernovae, and the internal structure of the Earth.

In order to resolve the mass hierarchy, detectors must determine the neutrino energy and arrival direction\footnote{Although only the neutrino zenith angle is relevant for the neutrino mass hierarchy measurement, the space angle for direction resolutions is always considered in this paper.} --- the latter defining the neutrino path through the Earth --- and identify the type of interaction, as electron-neutrino charged current (CC), muon-neutrino CC, or neutral current (NC).\footnote{At these energies, the much-rarer $\nuan_{\tau}$ interactions will resemble one of these classes.} With ORCA, this will be done by studying the Cherenkov light signature using an array of PMTs. A first estimate of the accuracy of reconstruction methods targeting the primary outgoing lepton ($e$ or $\mu$) in CC interactions is given in the Letter of Intent for KM3NeT 2.0~\cite{KM3NeT_LoI}.

Improving reconstruction methods will require using the hadronic component of CC interactions, in particular since it has been suggested to resolve the inelasticity $y$ of the interaction to further add to the mass hierarchy sensitivity~\cite{2013PhRvD..87k3007R}. However,
the instrumentation density of the ORCA detector will likely be insufficient to identify the Cherenkov cone of each individual particle (as with Super-Kamiokande~\cite{2003NIMPA.501..418F}), or measure the paths of particles as a tracking detector. Effects causing fluctuations in the Cherenkov light signature will therefore limit the achievable reconstruction accuracy.

In this paper, the limits imposed by these intrinsic physical fluctuations on the energy and direction resolution of hadronic cascades are derived (as per section~\ref{sec:methods}) by analysing the output of simulations for seawater in section~\ref{sec:hadronic_cascades}. Limits on resolutions of muon tracks and electromagnetic cascades are much smaller and hence less significant, and are derived in appendices~\ref{sec:muon_tracks} and~\ref{sec:electron_cascades}. Combining these fluctuations produces limiting resolutions for the neutrino in $\nuan_{e}$ and $\nuan_{\mu}$~CC interactions, as presented in section~\ref{sec:neutrinos}.\footnote{Since the mean inelasticity of $\nu$ and $\overline{\nu}$ interactions differ, throughout this paper the notation $\nuan$ refers to ``both $\nu$ and $\overline{\nu}$'', while $\nu$ and $\overline{\nu}$ refers exclusively to neutrinos and antineutrinos, respectively.} The effects of these limits on the reconstruction of the inelasticity in ORCA are investigated in section~\ref{sec:inelasticity}. Section~\ref{sec:discussion} compares the derived limits to the reconstruction performance of ORCA using a full detector simulation, discusses their applicability to reconstruction in ice-based detectors (i.e.\ PINGU), and their implications for published estimates of experimental sensitivity to the neutrino mass hierarchy.
The outcomes of this investigation are summarised in section~\ref{sec:conclusions}.

\section{Methods}
\label{sec:methods}

\subsection{Definitions}

\begin{figure*}[tbp]
\begin{center}
\includegraphics[width=0.5\textwidth]{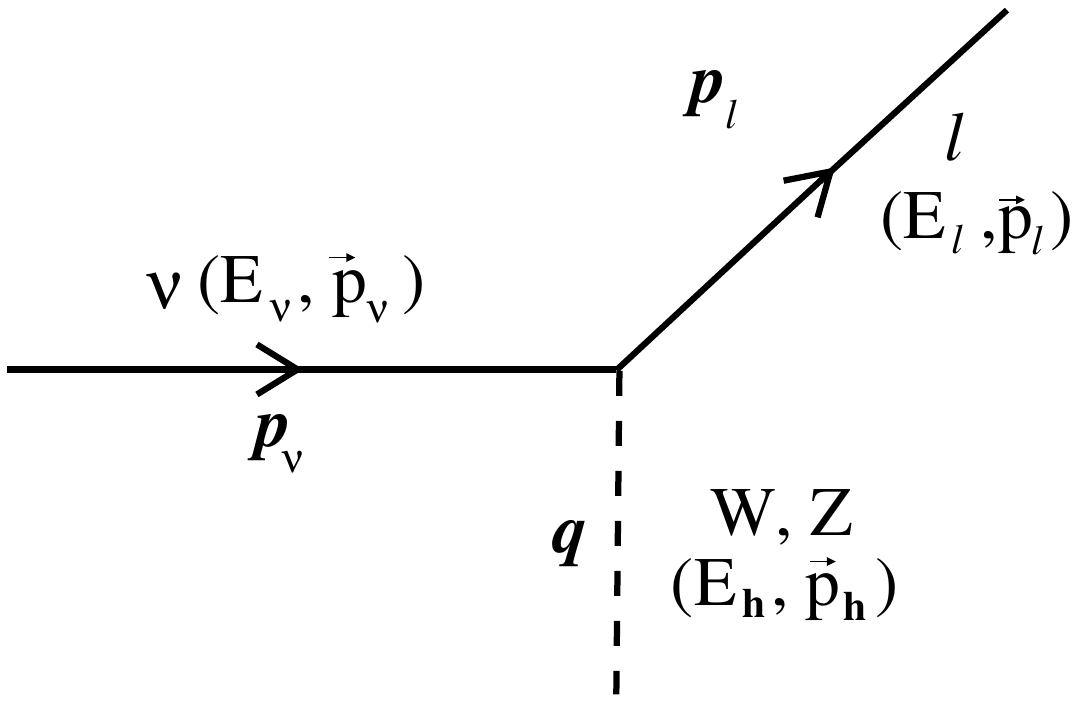}
\end{center}
\caption{
Sketch illustrating the definition of hadronic energy and momentum $\left(E_h, \vec{p}_h \right)$ used in this paper as the four-momentum transfer $\mathbf{q}$ between the incoming neutrino $\left(E_\nu, \vec{p}_\nu \right)$ and the outgoing lepton $\left(E_{\ell}, \vec{p}_{\ell} \right)$ in a neutrino interaction.
}
\label{fig:vertex}
\end{figure*}

With the exception of scattering on atomic electrons, all neutrino interactions in the GeV range produce hadronic cascades, although the `cascade' may only consist of a few particles emerging from the neutrino interaction. There are several possible definitions of the energy and momentum of such a hadronic cascade, $\mathbf{p}_h = \left(E_h, \vec{p}_h \right)$, which may or may not include the target momentum, any sub-relativistic nuclear remnant, and may also weight outgoing particles by their expected Cherenkov light yield. The convention chosen here is to define $\mathbf{p}_h$ as the momentum transfer $\mathbf{q}$, i.e.\ through four-momentum conservation applied to the neutrino interaction vertex in the laboratory frame, as shown in figure~\ref{fig:vertex}. Thus,
\begin{equation}
\mathbf{p}_h   \equiv  \mathbf{q}   =  \mathbf{p}_{\nu} - \mathbf{p}_{\ell} , \label{eq:ph}
\end{equation}
where $\mathbf{p}_{\nu}$ is the original neutrino four-momentum, and $\mathbf{p}_{\ell}$ the four-momentum of the outgoing lepton. Similarly, the inelasticity $y$ (nominally, Bjorken $y$) of the interaction is defined as 
\begin{equation}
y \, \equiv \, \frac{E_{h} }{E_{\nu}} \, = \, \frac{E_{\nu}-E_{\ell} }{E_{\nu}}.
\label{eq:y}
\end{equation}
These definitions are chosen because they are most relevant for reconstructing the neutrino interaction properties.  By not including the (unobservable) target four-momentum, variations in this contribution between interactions with the same $\mathbf{p}_h$ will be included in the fluctuations due to different hadronic final states.

In ORCA and like detectors, events are reconstructed by measuring the number of emitted photons, their arrival times and their spatial distribution. Unlike detectors using a dense array of PMTs on a $2$D plane, such as Super-Kamiokande~\cite{2003NIMPA.501..418F}, the sparser $3$D layout of ORCA makes it difficult to distinguish Cherenkov cones from individual particles. Therefore, it is assumed that signatures from the outgoing $e,\mu$ in $\nuan_{e,\mu}$~CC interactions can be distinguished from the accompanying hadronic cascade as demonstrated in ref.~\cite{KM3NeT_LoI}, but that individual particles within the hadronic cascade cannot be identified.

Moreover, while the elongation of the cascade is important for a dense detector configuration, effects due to this elongation are not studied here, and only fluctuations in the emission direction and number of photons are considered. Furthermore, it is assumed that only unscattered photons retain directional information. Therefore, the total number of detected photons can be used to estimate the cascade energy $E_h$, while the emitted directions of photons that are detected before being scattered can be used to estimate the cascade direction, $\vec{u}_{h}$ ($= \vec{p}_{h} / \left | \vec{p}_{h}  \right |$).

Fluctuations in reconstructed energy and direction are considered according to the following scheme:

\begin{itemize}
\item{\it Hadronic-state fluctuations.} Hadronic final states with identical $\mathbf{p}_h$ can differ in the number and type of particles, and in their energies and directions. This will result in different photon signatures, and hence fluctuations in reconstructed quantities.
\item{\it Propagation fluctuations.} The stochastic nature of energetic particle propagation implies that even identical initial particles will produce different photon signatures.
\item{\it All-photon limit.} Even in the case that every emitted photon is detected, 
the above two effects will combine to give the all-photon limit to reconstruction accuracy.
\item{\it Photon-sampling fluctuations.} Sampling only a fraction of the emitted photons will further limit the ability of reconstruction methods to determine event properties.
\item{\it Overall limit.} The combination of hadronic-state, propagation, and photon-sampling effects results in the overall limit to reconstruction accuracy.
\end{itemize}

Values of the relative energy reconstruction error, $\Delta E/E$, and the direction reconstruction error, $\Delta \theta$, will be derived for each of these cases.

\begin{figure*}[tbp]
\begin{center}
\includegraphics[width=0.8\textwidth]{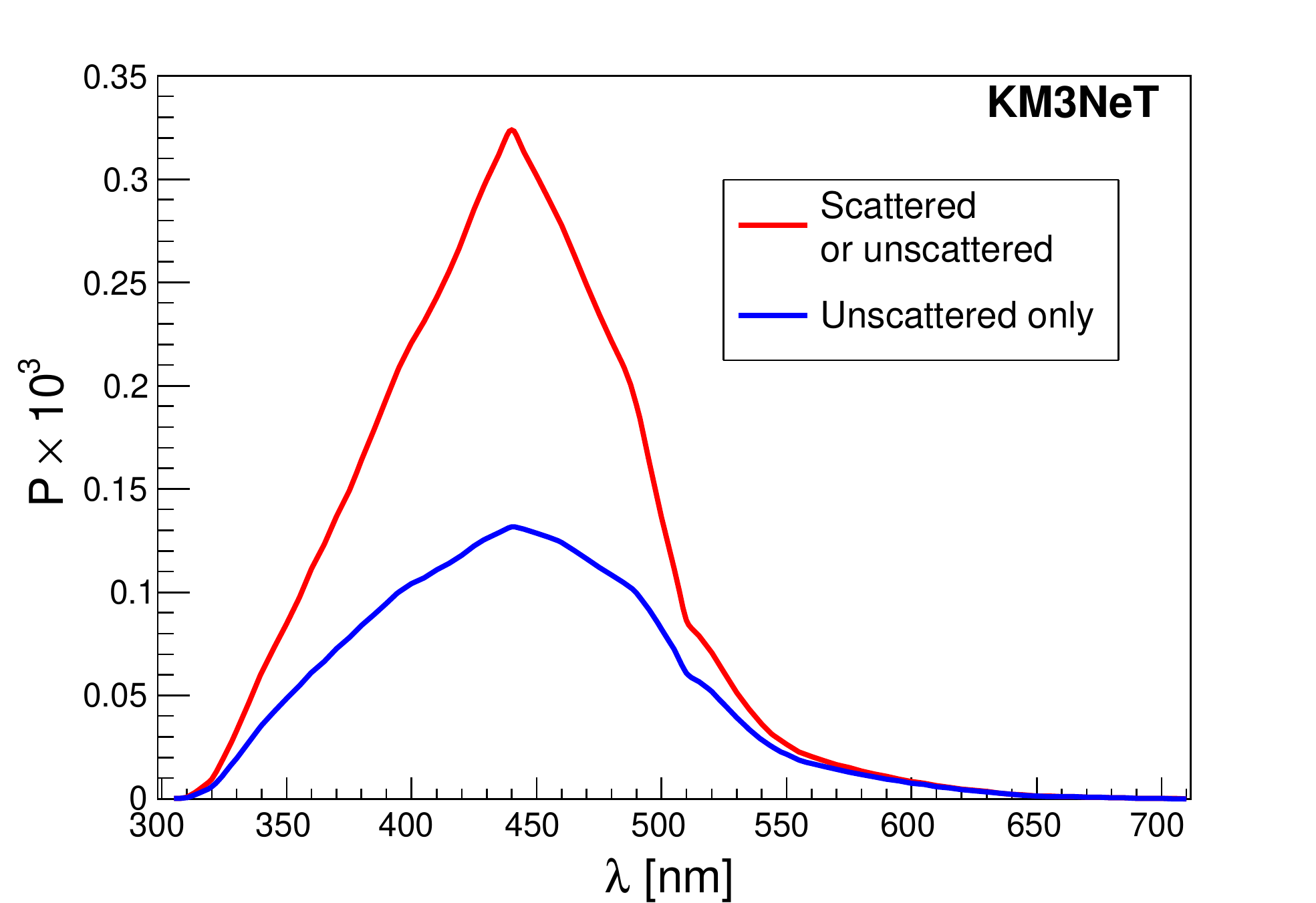}
\end{center}
\caption{
Detection probability, $P$, for all unabsorbed photons (red) and for those that are additionally not scattered (blue), calculated from in-situ measurements of the inherent optical properties of deep-sea water~\cite{2007APh....27....1R}, and the optical module effective area and density for the KM3NeT/ORCA benchmark detector design~\cite{KM3NeT_LoI}. The peak near $440$\,nm arises from the absorption minimum.}
\label{fig:p_det_dir}
\end{figure*}

The total and unscattered sampled photon fractions are characterised only by the wavelength-dependent probability of detecting each photon, given in figure~\ref{fig:p_det_dir} for the ORCA benchmark detector studied in ref.~\cite{KM3NeT_LoI}. Note that this detector has a $6$\,m vertical spacing between optical modules, while a $9$\,m spacing was found to be optimal for the neutrino mass hierarchy determination.

As all sources of fluctuations are independent, the all-photon and overall limits are calculated by adding their component contributions in quadrature.

\subsection{Simulations}
\label{sec:simulations}

The simulation methods used in this paper to calculate limiting resolutions differ significantly from reconstruction methods based on full detector simulations. As the goal is to derive limiting resolutions in a robust manner, the following simplifications are used when compared to full simulations of ORCA~\cite{KM3NeT_LoI}. In all cases, these are intended to be optimistic, to ensure that the resulting limits are indeed limiting.

\begin{itemize}

\item{\it PMT response ignored.} The precise number and arrival times of photons are assumed to be measured. This ignores the specifics of PMT response and readout, timing- and charge-calibration uncertainties, and imperfections in reconstruction methods.

\item{\it Generalised detector geometry.} Detected photons are randomly sampled according to wavelength-dependent probabilities only. This ignores the specifics of the detection geometry, such as partially contained events, and that PMTs are located in `clumps' (i.e.\ optical modules) at specific positions. This simplified geometry is equivalent to a detector composed of an infinite number of infinitely small detection elements, and thus ignores the non-uniform and correlated photon-detection induced by `clumpiness'.

\item{\it Known photon origin.} Photons from the hadronic cascades and the outgoing lepton in a CC interactions are assumed to be distinguishable, and background photons from atmospheric muons and natural light sources are ignored.

\item{\it No simulation uncertainties.} It is assumed that neutrino interactions, particle propagation, and Cherenkov light production can be modelled perfectly, and reconstructions are based on perfect simulations.

\end{itemize}

Initial hadronic states are taken from simulations of neutrinos interacting in seawater using {\tt gSeaGen}~\cite{vlvnt_gseagen}, which implements {\tt GENIE 2.8.4}~\cite{2010NIMPA.614...87A}. Samples of $1000$ such states for each of several discrete energies between $1$\,GeV and $20$\,GeV were selected in a sufficiently unbiased manner (see appendix~\ref{sec:appendix_selecting_had} and table~\ref{tab:simulations} for further details). Variations about the mean properties of each sample characterise the hadronic-state fluctuations.

Each state was simulated $1000$ times with an implementation of {\tt GEANT 3.21}~\cite{geant3} in the standard ANTARES simulation package~\cite{2013NIMPA.725...98M}. Cherenkov emission was generated over the $300$--$710$\,nm range
via the Frank-Tamm formula~\cite{FrankTamm37} using the wavelength-dependent phase velocity in seawater~\cite{2005APh....23..131A}.

Variations between simulations of the same cascade are used to characterise fluctuations due to propagation. Implementations of both {\tt GHEISHA} (`{\tt G-GHEISHA}')~\cite{gheisha} and {\tt FLUKA} (`{\tt G-FLUKA}')~\cite{fluka2005,2014NDS...120..211B} were used for hadron tracking, with differences (discussed in appendix~\ref{sec:appendix_GHEISHA_vs_FLUKA}) serving as a measure of systematic uncertainties. 
In most cases, this difference was found to be small. Neutrino resolutions presented in section~\ref{sec:neutrinos} use {\tt G-GHEISHA} results.

For each simulation of each hadronic final state, the properties of all emitted photons, and a sample randomly selected according to the detection probabilities of figure~\ref{fig:p_det_dir}, are recorded, and used to estimate corresponding reconstruction errors as described below. Distributions of parameters related to reconstructed energy are approximately Gaussian, and fluctuations are characterised by their root-mean-square (RMS). Those related to direction reconstruction tend to have large tails due to individually significant scattering events, and are characterised using $68.27$\% quantiles in space angle.

In case of neutrino resolutions, the Monte Carlo truth values of $\mathbf{p}_l$ and $\mathbf{p}_h$ are taken from {\tt gSeaGen}. The errors in reconstructing them are calculated by considering the leptonic and hadronic components separately, i.e.\ it is assumed that the photons from the lepton can be identified. Correlations between energy and direction errors for each component are accounted for. A brief investigation has shown correlations with other properties of neutrino interactions to be very small (appendix~\ref{sec:appendix_selecting_had}), so they are ignored in this work.

A further approximation is required when $E_h$ or $E_{e,\mu}$ do not equal one of the discrete simulated energies. In such cases, the expected properties of $\Delta E/E$ and $\Delta \theta$ are interpolated for $E > 1$\,GeV, and extrapolated for $E < 1$\,GeV. Thus, neutrino resolution limits depending on components --- especially hadronic cascades --- with $E < 1$\,GeV should be interpreted with care.

\section{Hadronic cascade reconstruction}
\label{sec:hadronic_cascades}

\subsection{Energy resolution}
\label{sec:hadron_energy}

\begin{figure*}[tbp]
\begin{center}
\includegraphics[width=0.8\textwidth]{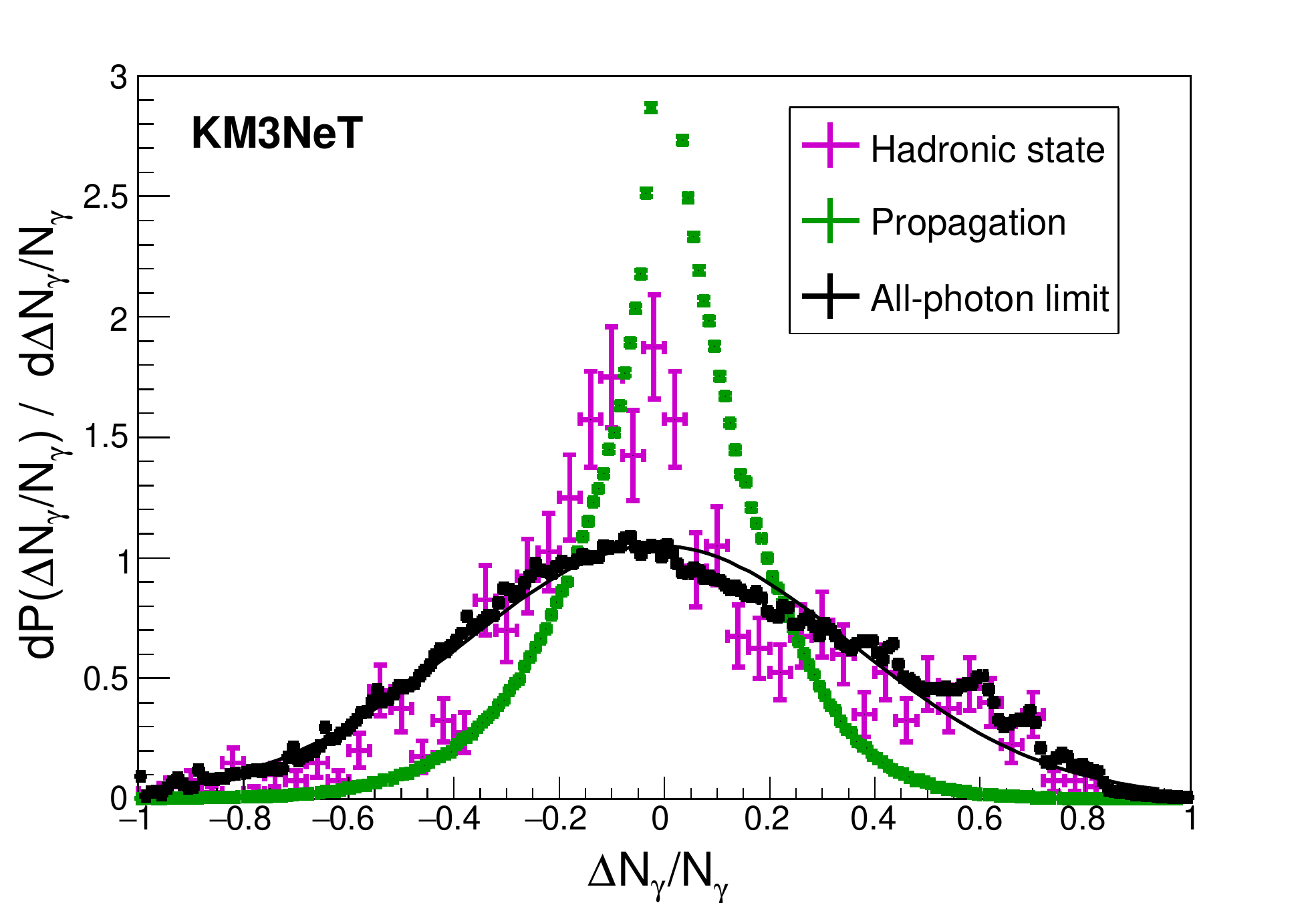}
\end{center}
\caption{
Probability density distributions of the relative deviations from the mean emitted photon number, $\Delta N_{\gamma}/N_{\gamma}$, in hadronic cascades with $E_h = 5$\,GeV. Shown are the contributions from hadronic state (purple), particle propagation (green), and their combined effect in the all-photon limit (black).
A Gaussian fit (black line) to the all-photon limit is shown to illustrate its appropriateness in characterising these fluctuations.}
\label{fig:hadronic_5gev_energy}
\end{figure*}

An example of the fluctuations in the photon yield for a hadronic cascade with $E_h=5$\,GeV is given in figure~\ref{fig:hadronic_5gev_energy}. The effect of different hadronic states is asymmetric, with relatively many events producing a large number of photons. A brief investigation suggests that these are hadronic cascades with a large electromagnetic component ($\gamma$, $e^{\pm}$, $\pi^0$). Particle propagation effects however are more symmetric and centrally-peaked. Their combined effect in the all-photon limit is compared to a Gaussian fit.

\begin{figure*}[tbp]
\begin{center}
\includegraphics[width=0.8\textwidth]{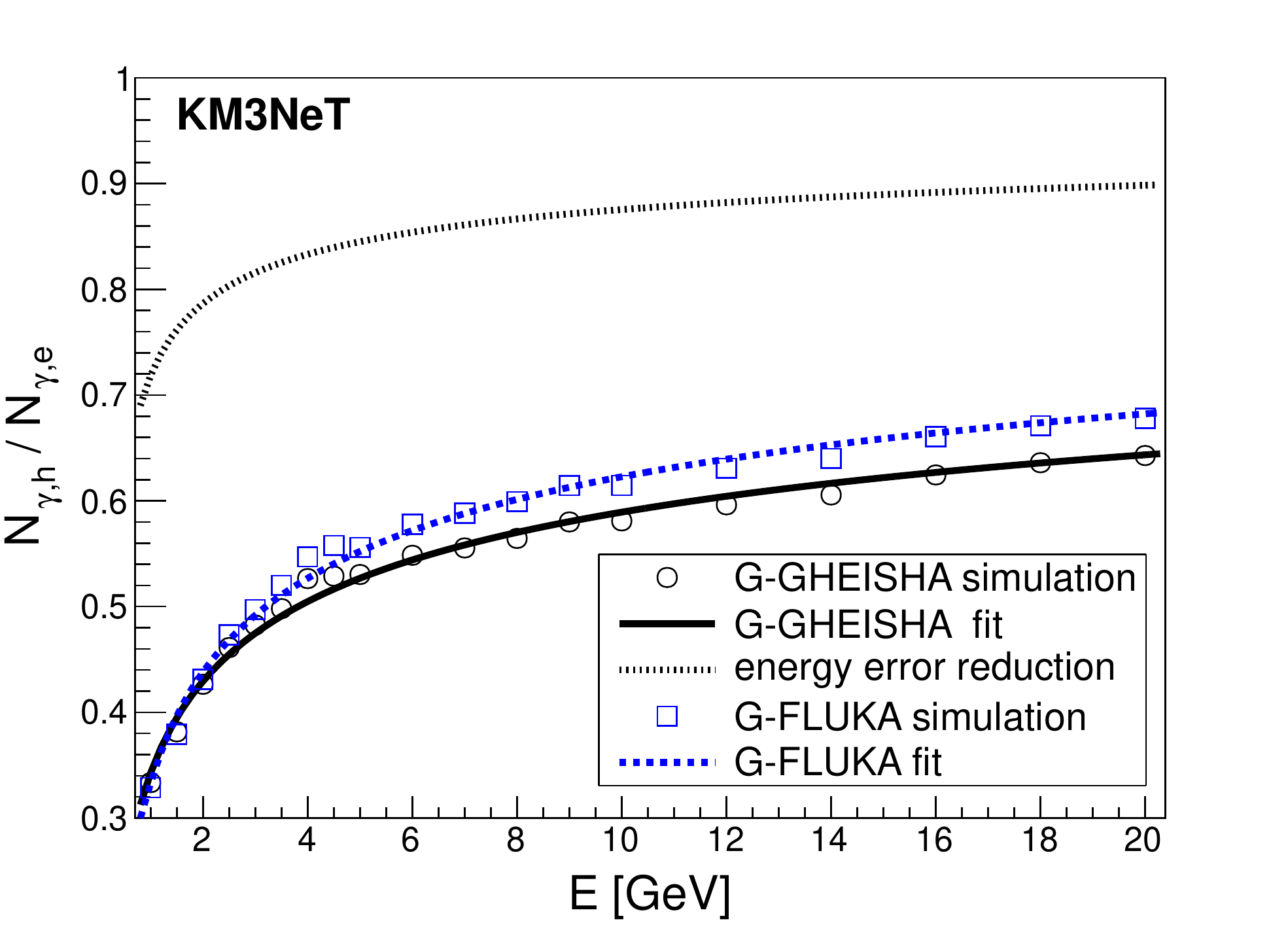}
\end{center}
\caption{
Number of photons emitted in the case of a hadronic cascade $N_{\gamma,h}$ relative to an electromagnetic cascade $N_{\gamma,e}$ of the same energy, showing data points and the fitted function (eq.~\ref{eq:LYratio}). Cascades simulated with {\tt G-GHEISHA} ({\tt G-FLUKA}) are shown in black (blue). Also shown for {\tt G-GHEISHA} is the resulting error reduction factor as defined in eq.~\ref{eq:correction} (black dotted line).
} \label{fig:fit_fh}
\end{figure*}

The hadronic cascade energy $E_h$ can be estimated from the total number of detected photons, with fluctuations therein producing fluctuations in the estimated energy. Since the total number of emitted photons is not linearly proportional to hadronic cascade energy (as in the case of electrons), the hadronic cascade energy is estimated using a fit to the total number of photons, $N_{\gamma,h}$. This can be expressed as a fraction $f_h$ relative to that from electromagnetic cascades of the same energy, $N_{\gamma,e}$ (appendix~\ref{sec:electron_cascades}), i.e.\ $N_{\gamma,h}(E_h) = f_h(E_h)\, N_{\gamma,e}(E_h)$. The fit,
\begin{eqnarray}
f_h(E_h) & = & 1  - 0.681 \left( \frac{E_h}{0.863\,\text{GeV}} \right)^{-0.207},
\label{eq:LYratio}
\end{eqnarray}
uses the same functional form as ref.~\cite{Kowalski:2004qc}, and is shown in figure~\ref{fig:fit_fh} (black solid line). The energy dependence of $f_h$ means that a deviation in the number of photons will have a non-linear effect on the reconstructed energy. The relative error in reconstructed energy can be calculated via
\begin{eqnarray}
\frac{\Delta E}{E} & \approx & \frac{N_{\gamma}}{E} \frac{d E}{d N_{\gamma}} \frac{\Delta N_{\gamma}}{N_{\gamma}}
=  f_h \, \frac{d E}{d (E_h f_h)} \, \frac{\Delta N_{\gamma}}{N_{\gamma}}, \label{eq:correction}
\end{eqnarray}
where the coefficient of $\Delta N_{\gamma}/N_{\gamma}$ is the energy error reduction factor and is also shown in figure~\ref{fig:fit_fh}. As $f_h$ is an increasing function, the resulting relative errors in reconstructed energy tend to be smaller than the intrinsic relative $N_\gamma$ fluctuations. For hadronic state and propagation fluctuations, variation in the Monte Carlo values of all $N_\gamma$ about the mean value are used for $\Delta N_{\gamma}$, while for photon sampling, the Poisson variation in the number of detected photons, $N_{\gamma}^{\rm det}$, is used:
\begin{eqnarray}
\left(\frac{\Delta N_{\gamma}}{N_{\gamma}} \right)^{\rm sampling} & = & \left(N_{\gamma}^{\rm det} \right)^{-0.5}.
\label{eq:poisson_var}
\end{eqnarray}

\begin{figure*}[tbp]
\begin{center}
\includegraphics[width=0.8\textwidth]{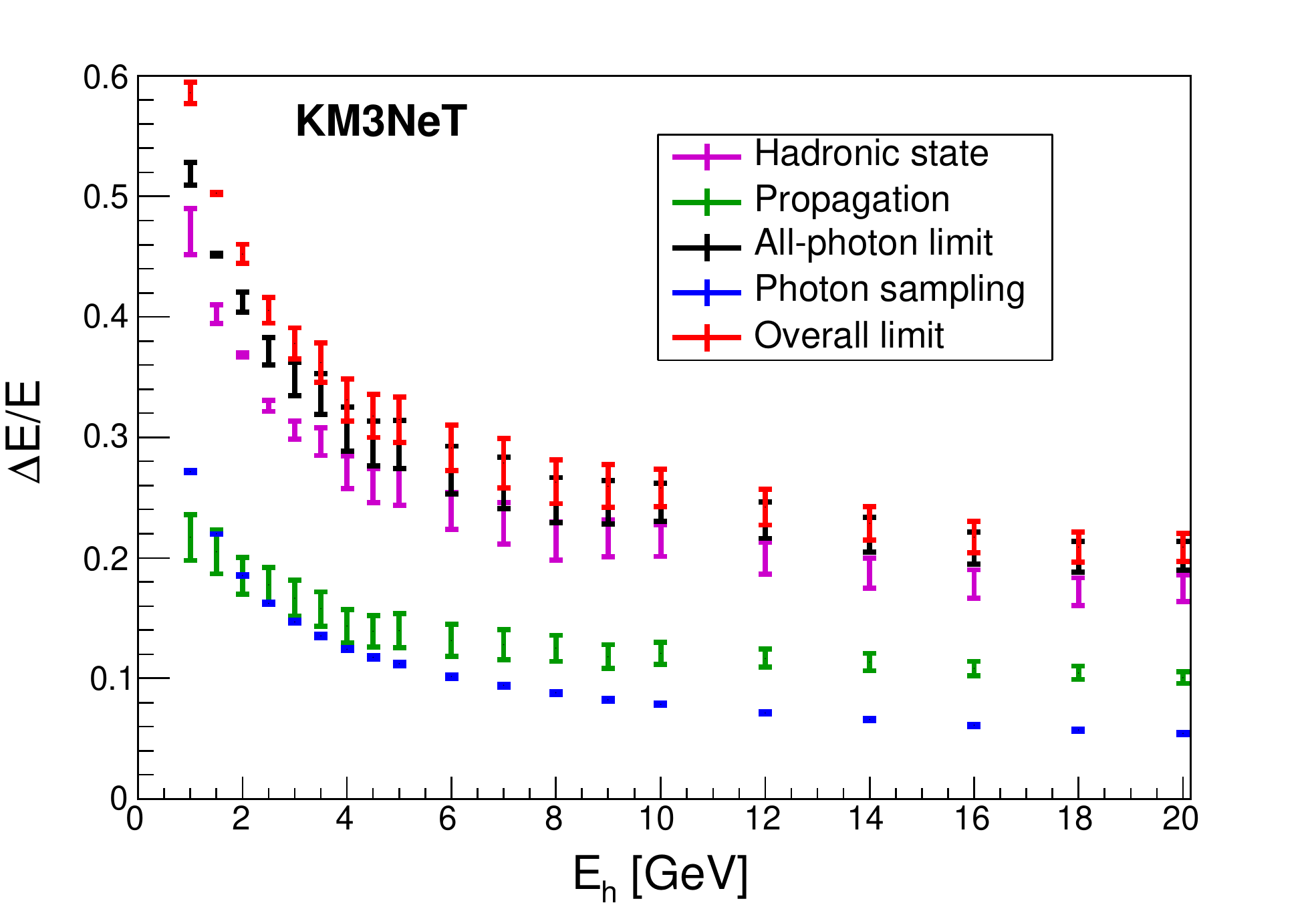}
\end{center}
\caption{
Relative errors (RMS) in reconstructed energy for hadronic cascades, due to hadronic state (purple), particle propagation (green), their combined effect in the all-photon limit (black), and the additional variation introduced due to photon sampling (blue) in the overall limit (red). The mean is calculated from the average between the {\tt G-GHEISHA} (larger) and {\tt G-FLUKA} (smaller) results, and the error bars cover the range between them.
} \label{fig:hadronic_energy_results}
\end{figure*}

The resulting relative energy errors, calculated by applying eq.~\ref{eq:correction} to relative variations in photon number from simulations (e.g.\ figure~\ref{fig:hadronic_5gev_energy}), are shown in figure~\ref{fig:hadronic_energy_results}. In the all-photon limit, fluctuations in hadronic state dominate over the whole energy range. The effect due to photon sampling is relatively small, so that the overall limiting resolution is only slightly larger than the all-photon limit.

The light yield of hadronic cascades (figure~\ref{fig:fit_fh}) and its intrinsic fluctuations (figure~\ref{fig:hadronic_energy_results}) were also investigated by ref.~\cite{Kowalski:2004qc} down to 10\,GeV, with broadly similar results.

\subsection{Direction resolution}
\label{sec:hadron_direction}

\begin{figure*}[tbp]
\begin{center}
\includegraphics[width=0.8\textwidth]{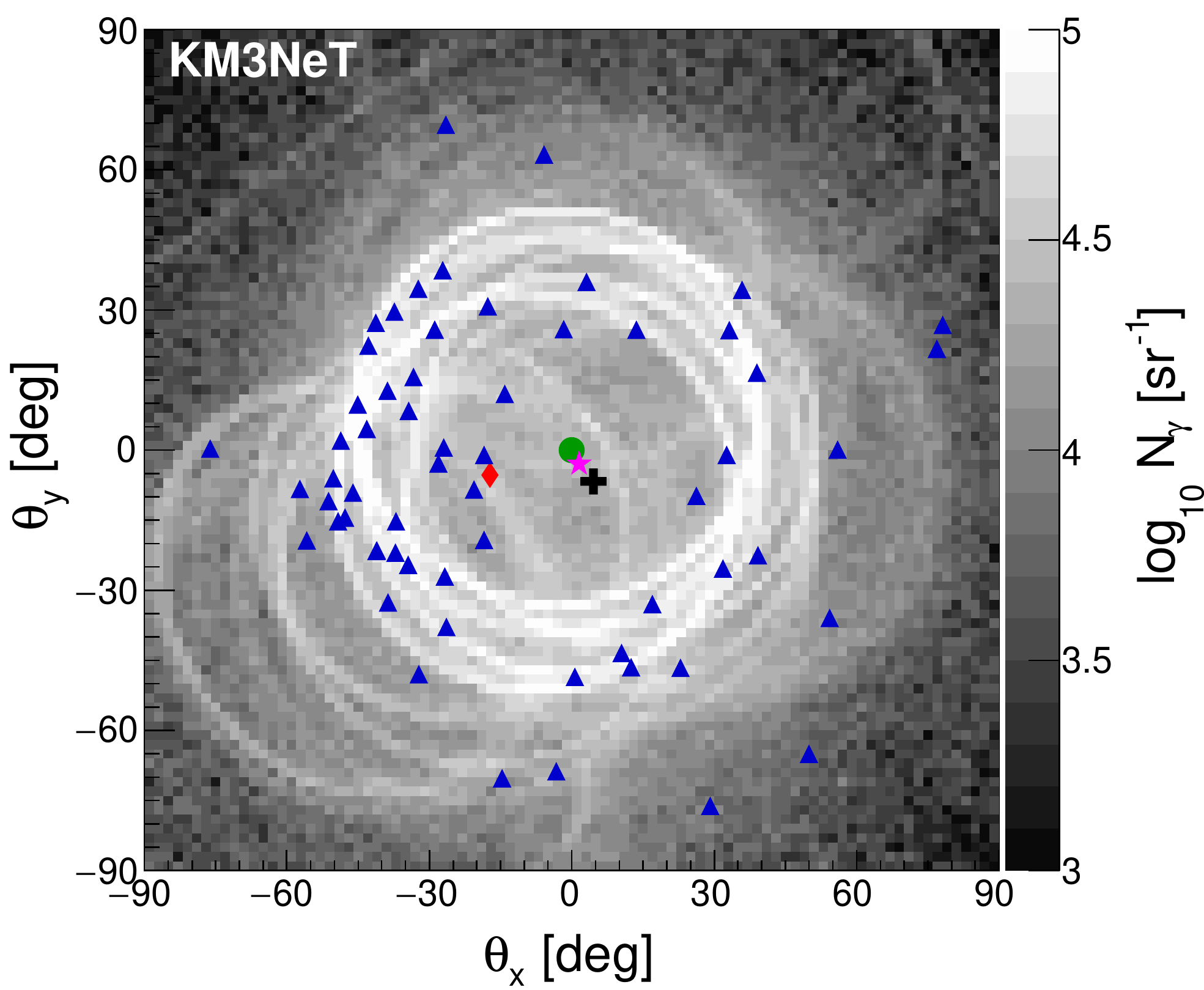}
\end{center}
\caption{Photon distributions from a hadronic cascade with $E_h=10$\,GeV. The true cascade direction is in the $\vec{u}_z$ direction, at $(\theta_x , \theta_y )= (0,0)$ (green circle). The mean photon direction, averaged over $1000$ simulation iterations with the same hadronic final state, is given by the purple star --- the $3^{\circ}$ offset from $(0,0)$ reflects the intrinsic bias for this particular state.
The distribution of all emitted photons for one particular propagation iteration is shown by the grey shading. The mean direction of all these photons is given by the black plus sign --- the $8^{\circ}$ offset of this from $(0,0)$ measures the total intrinsic variation in the case of the all-photon limit, while the $5^{\circ}$ offset from the purple star measures the variation due to random particle propagation only. A random sample of unscattered photons, chosen according to the probabilities of figure~\ref{fig:p_det_dir}, are shown by blue triangles, along with their mean direction (red diamond). The $25^{\circ}$ offset of the red diamond from the black cross measures the variation due to photon sampling only, while the offset of $18^{\circ}$ from $(0,0)$ gives the overall variation.
} \label{fig:hadr_angle_method}
\end{figure*}

\begin{figure*}[tbp]
\begin{center}
\includegraphics[width=0.8\textwidth]{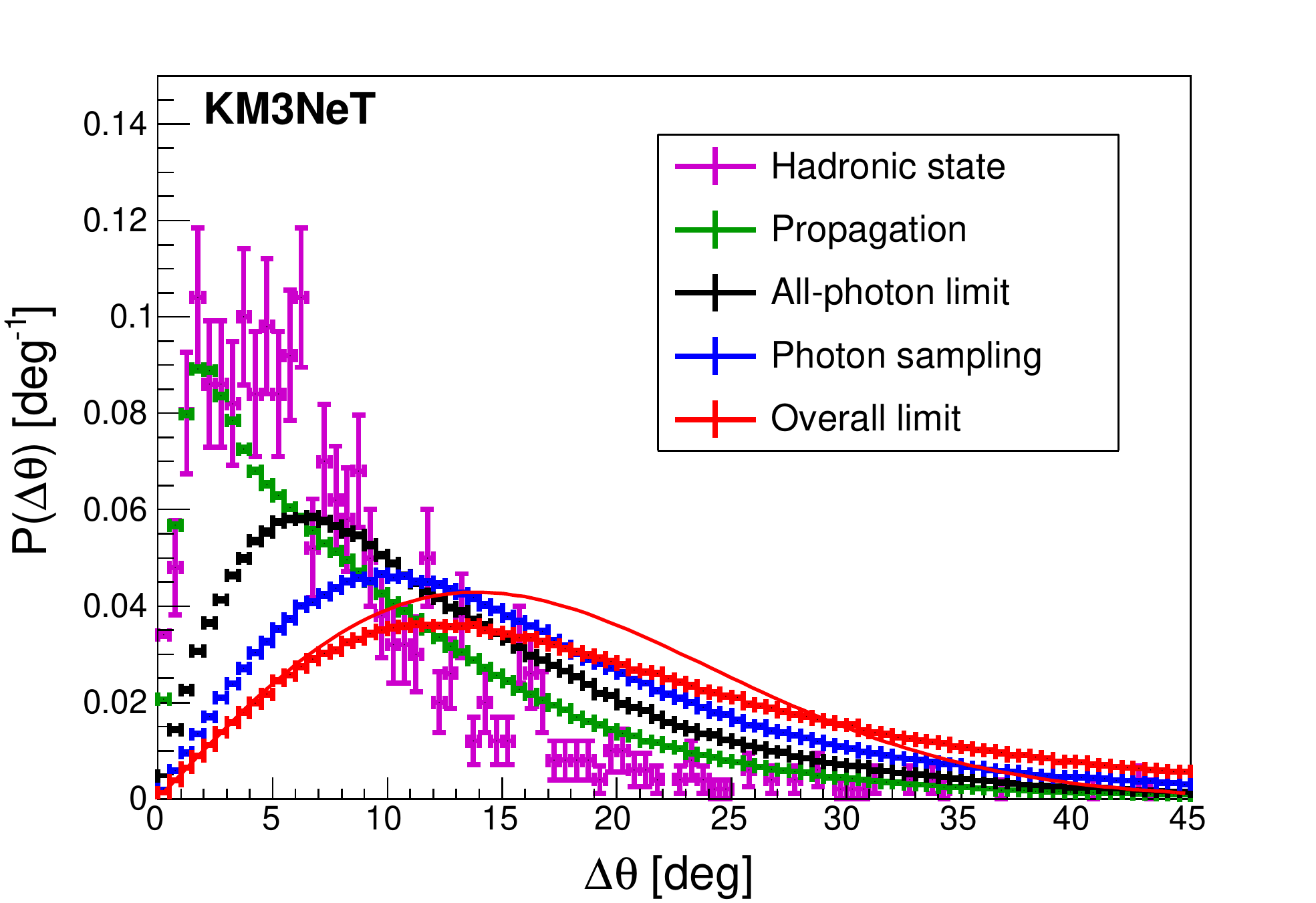}
\end{center}
\caption{
Distribution of direction errors $\Delta \theta$ for $E_h = 5$\,GeV, showing histograms from $1000$ simulations over each of $1000$ interactions. The three different sources of variation (hadronic state, propagation, and photon sampling) and their combined effects for all-photon and overall limits are shown in points, with statistical error bars.
A fit (red line) to the overall limit using the radial distribution function of a 2D-Gaussian is shown to illustrate its non-Gaussianity.
}
\label{fig:hadronic_5gev_direction}
\end{figure*}

A robust method to estimate the mean hadronic cascade direction is to use the mean direction of detected unscattered photons --- the validity of this method is discussed in appendix~\ref{sec:electron_direction}. In the all-photon limit, this estimate will be made using all emitted photons, while for photon sampling, only the detected unscattered fraction is used.  An illustration of the resulting effects on the direction resolution is given in figure~\ref{fig:hadr_angle_method} for a single hadronic final state. The distributions of direction errors for $E_h = 5$\,GeV are illustrated in figure~\ref{fig:hadronic_5gev_direction}.

\begin{figure*}[tbp]
\begin{center}
\includegraphics[width=0.8\textwidth]{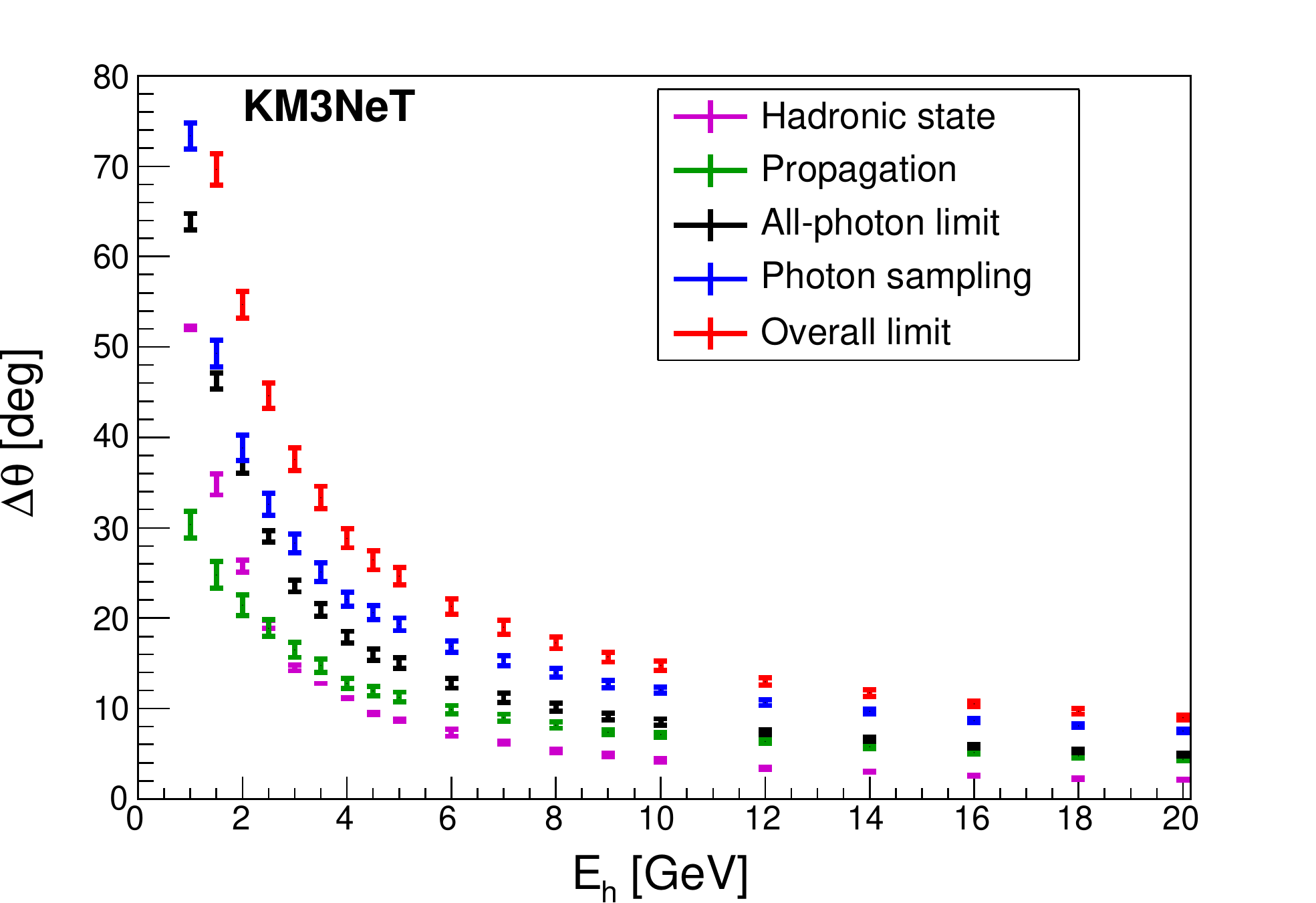}
\end{center}
\caption{
Direction errors $\Delta \theta$ of hadronic cascades (68\% quantiles). For explanations, see figure~\ref{fig:hadronic_energy_results}.}
 \label{fig:hadronic_direction_results}
\end{figure*}

Limiting direction resolutions for hadronic cascades between $1$\,GeV and $20$\,GeV are presented in figure~\ref{fig:hadronic_direction_results}. Below $\sim 2$\,GeV, variation in the initial hadronic state dominates the direction resolution in the all-photon limit, while above this energy particle propagation effects dominate. Over the entire energy range, their combined effect in the all-photon limit is less significant than the effects of photon sampling, so that the overall limiting resolution is worse. The relative difference in results between the {\tt G-FLUKA} and {\tt G-GHEISHA} cases is less than $\sim 10$\%.

\section{Resolutions of muon- and electron-neutrino charged-current events}
\label{sec:neutrinos}

Using the results for hadronic cascades of the previous section, and adding results for muon tracks and electron cascades from appendices~\ref{sec:muon_tracks} and~\ref{sec:electron_cascades}, limits on the reconstruction accuracy of $\nuan_{\mu}$ and $\nuan_e$~CC events in the $1$--$20$\,GeV range can be deduced.\footnote{The accuracy for NC events of all flavours can be deduced from the interaction kinematics and the limiting resolutions for hadronic cascade reconstruction (section~\ref{sec:hadronic_cascades}), while the more complex, and much rarer, $\nuan_\tau$~CC events can be approximated by combining electron, muon, and hadronic cascade resolutions according to $\tau$ decay products.}

\begin{figure*}[tbp]
\begin{center}
\includegraphics[width=0.8\textwidth]{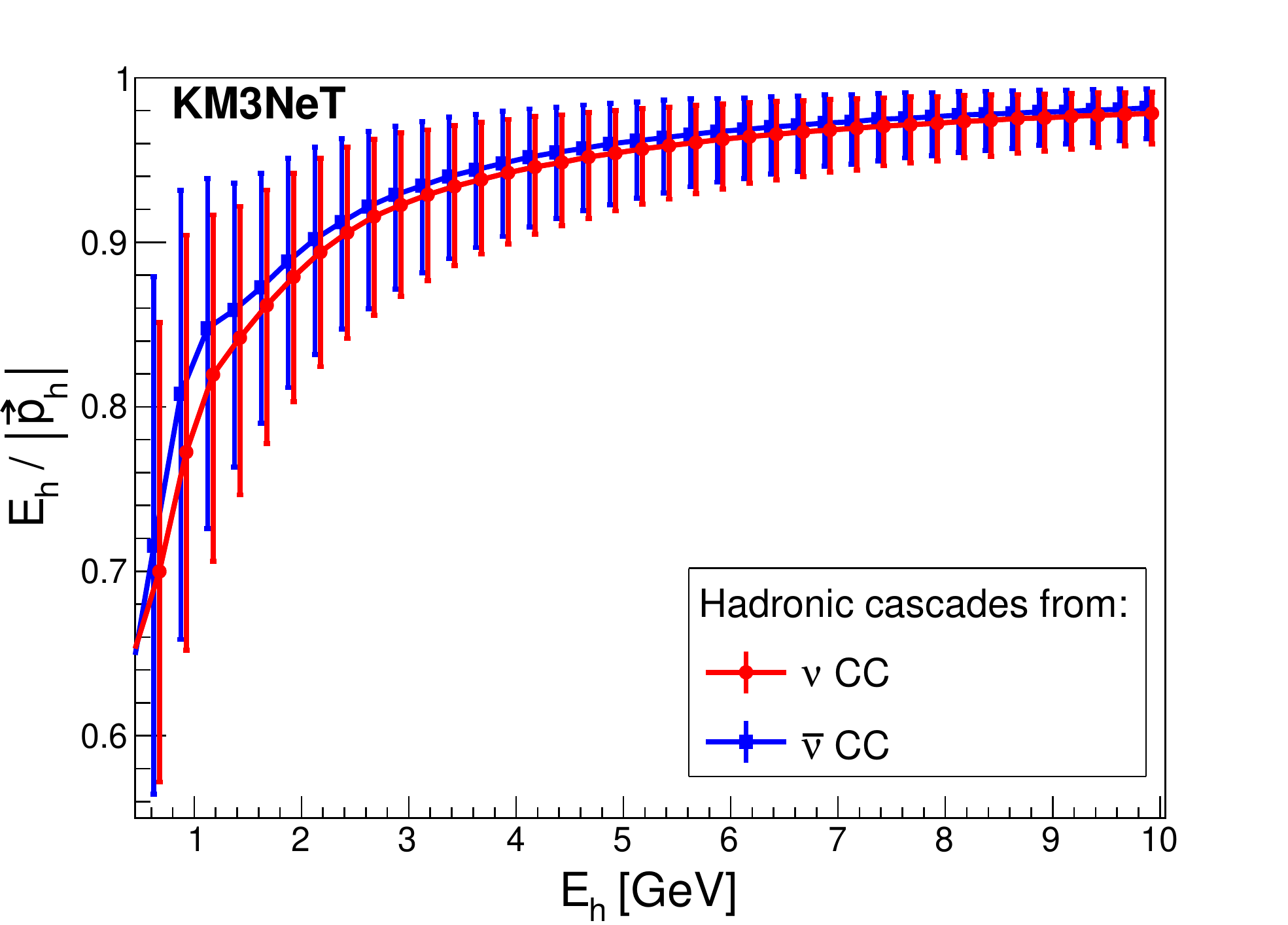}
\end{center}
\caption{Ratio $E_h/|\vec{p}_h|$ (calculated according to eq.~\ref{eq:ph}) as a function of $E_h$, for hadronic cascades extracted from both $\nu$~CC (red) and $\overline{\nu}$~CC (blue) events. Error bars represent the variation characterised by the RMS. Note that the definition of $\mathbf{p}_h \equiv \mathbf{q}$ (eq.~\ref{eq:ph}, and figure~\ref{fig:vertex}) requires $|\vec{p}_h| > E_h$, since $q^2$ is negative.}
\label{fig:e_p_fig}
\end{figure*}

The neutrino energy $E_\nu$ and normalised direction, $\vec{u}_\nu$, can be reconstructed using the interaction kinematics of eq.~\ref{eq:ph}, i.e.\ by combining the reconstructed lepton properties with that of the hadronic cascade:
\begin{eqnarray}
E_\nu^{\rm reco} & = & E_{\ell}^{\rm reco} + E_h^{\rm reco}, \nonumber\\
\vec{u}_\nu^{\rm reco} & = &  w_{\ell} \frac{E_{\ell}^{\rm reco}}{E_\nu^{\rm reco}} \, \vec{u}_{\ell}^{\rm reco} + w_h \frac{E_{h}^{\rm reco}}{E_\nu^{\rm reco}} \, \vec{u}_{h}^{\rm reco}.
\label{eq:nu_reco_Ereco_weighing}
\end{eqnarray}
Naively, the weights $w_\ell$ and $w_h$ could be set to unity, at which point eq.~\ref{eq:nu_reco_Ereco_weighing} closely resembles eq.~\ref{eq:ph}. However, there are several reasons for having $w_\ell$ and $w_h$ differ from unity. Firstly, while the hadronic cascade momentum magnitude $|\vec{p}_h|$ is assumed not to be reconstructable, it can be estimated using the reconstructed value of $E_h$ and the expected ratio $E_h/|\vec{p}_h|$, which is plotted in figure~\ref{fig:e_p_fig}. Fluctuations in this ratio can be considered as an additional hadronic-state contribution to neutrino direction resolution, which is not included in either the hadronic or lepton components.

A second factor influencing the weight is the relative accuracy of $\vec{u}_{\ell}$ and $\vec{u}_{h}$. In general, the best estimator of $\vec{u}_\nu$ will be found by reducing $w_h$ and increasing $w_\ell$ to reduce the influence of the much larger fluctuations in $\vec{u}_{h}$, even if this biases the resulting estimate in the direction of the lepton.

A third and final factor is the use of eq.~\ref{eq:ph} to constrain the measured values of $\mathbf{p}_h$ and $\mathbf{p}_{\ell}$, both by requiring they add to a physical value of $\mathbf{p}_{\nu}$, and by imposing prior likelihoods on the implied interaction kinematics.

\begin{figure*}[tbp]
\begin{center}
\includegraphics[width=0.8\textwidth, trim={0 5cm 0 0}, clip=true]{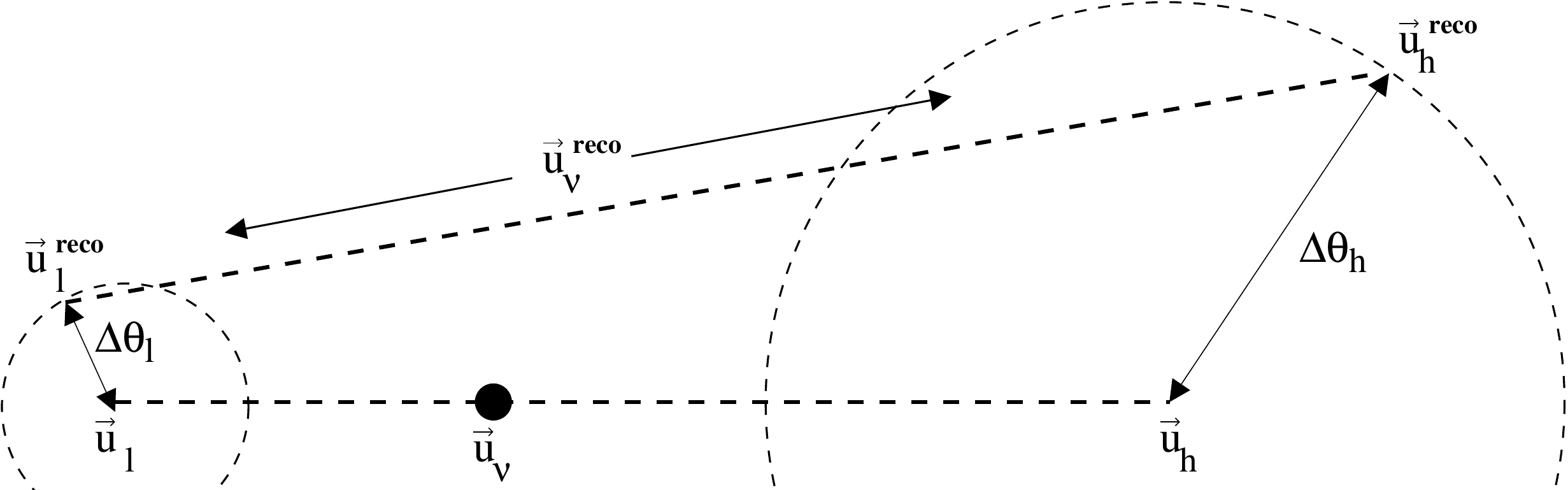}
\end{center}
\caption{Sketch illustrating the reconstruction of a neutrino direction $\vec{u}_{\nu}$. The true lepton and hadronic directions, $\vec{u}_l$ and $\vec{u}_h$, are reconstructed in directions $\vec{u}_l^{\rm reco}$ and $\vec{u}_h^{\rm reco}$ with characteristic errors $\Delta \theta_l$ and $\Delta \theta_h$. The reconstructed neutrino direction, $\vec{u}_{\nu}^{\rm reco}$, must then lie on the vector defined by these points. Sections~\ref{sec:reco_energy_momentum_conservation} and~\ref{sec:reco_expected_scattering_angle} essentially differ in the method for choosing a point on this vector.
}
\label{fig:lep_had_plane_sketch}
\end{figure*}

The general situation of choosing arbitrary weights between $\vec{u}_{\ell}$ and $\vec{u}_{h}$ is illustrated in figure~\ref{fig:lep_had_plane_sketch}. Ultimately, the optimum method would define weights based on the reconstructed energies of both components, and the angle between them, to account for the three aforementioned factors. However, such an optimisation is beyond the scope of this paper.

In the following, three methods are investigated. The first method (section~\ref{sec:reco_energy_momentum_conservation}) is the naive `four-momentum conservation' approach using $w_h = w_{\ell} = 1$. This is motivated by the momentum-magnitude and statistical down-weighting factors mentioned above tending to cancel each other, and because it is simple and robust. The second method (section~\ref{sec:reco_expected_scattering_angle}) accounts for physical constraints by using the expected scattering angle between the neutrino and lepton. The third method (section~\ref{sec:resolutions_elecCC_noHitID}) investigates the loss of accuracy when the electron and hadronic cascades in $\nuan_e$~CC events cannot be distinguished. In all cases, the `track length' method for energy reconstruction of muons is used (appendix~\ref{sec:muon_tracks}).

\subsection{Using naive four-momentum conservation}
\label{sec:reco_energy_momentum_conservation}

\begin{figure}[tbp]
\begin{center}
\includegraphics[width=0.8\textwidth]{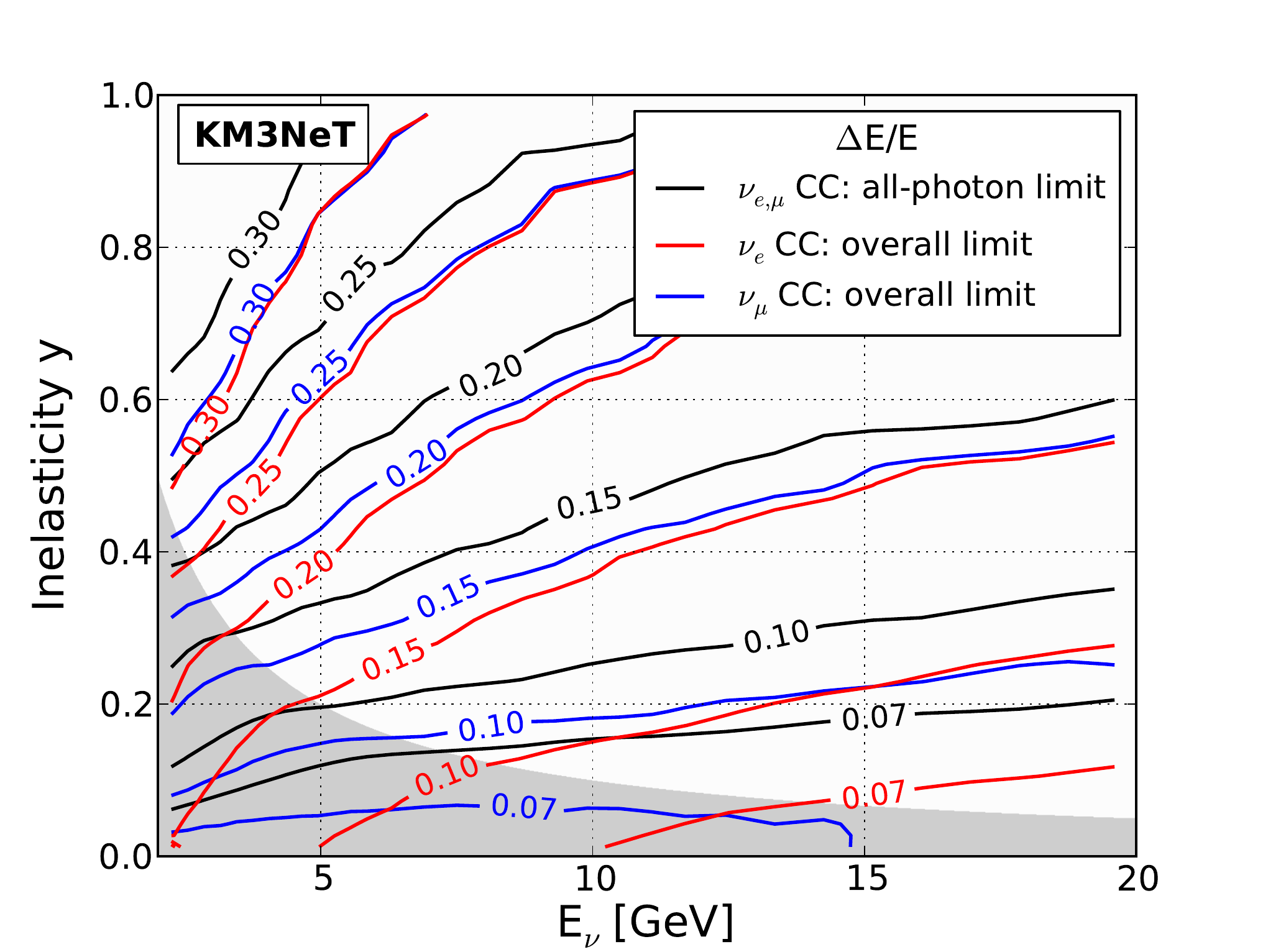}
\end{center}
\caption{Limitations on relative neutrino energy resolution (RMS) as a function of neutrino energy $E_\nu$ and inelasticity $y$. Resolutions are shown as contour lines for $\protect\nuan_{e,\mu}$~CC in the all-photon limit (black), for $\protect\nuan_{e}$~CC in the overall limit (red), and for $\protect\nuan_{\mu}$~CC in the overall limit (blue). The region $E_h < 1$\,GeV (grey shading) has been calculated using extrapolated hadronic cascade results, and should be interpreted with care.
}
\label{fig:reso2D_energy_numuANDnue}
\end{figure}
\begin{figure}[tbp]
\begin{center}
\includegraphics[width=0.8\textwidth]{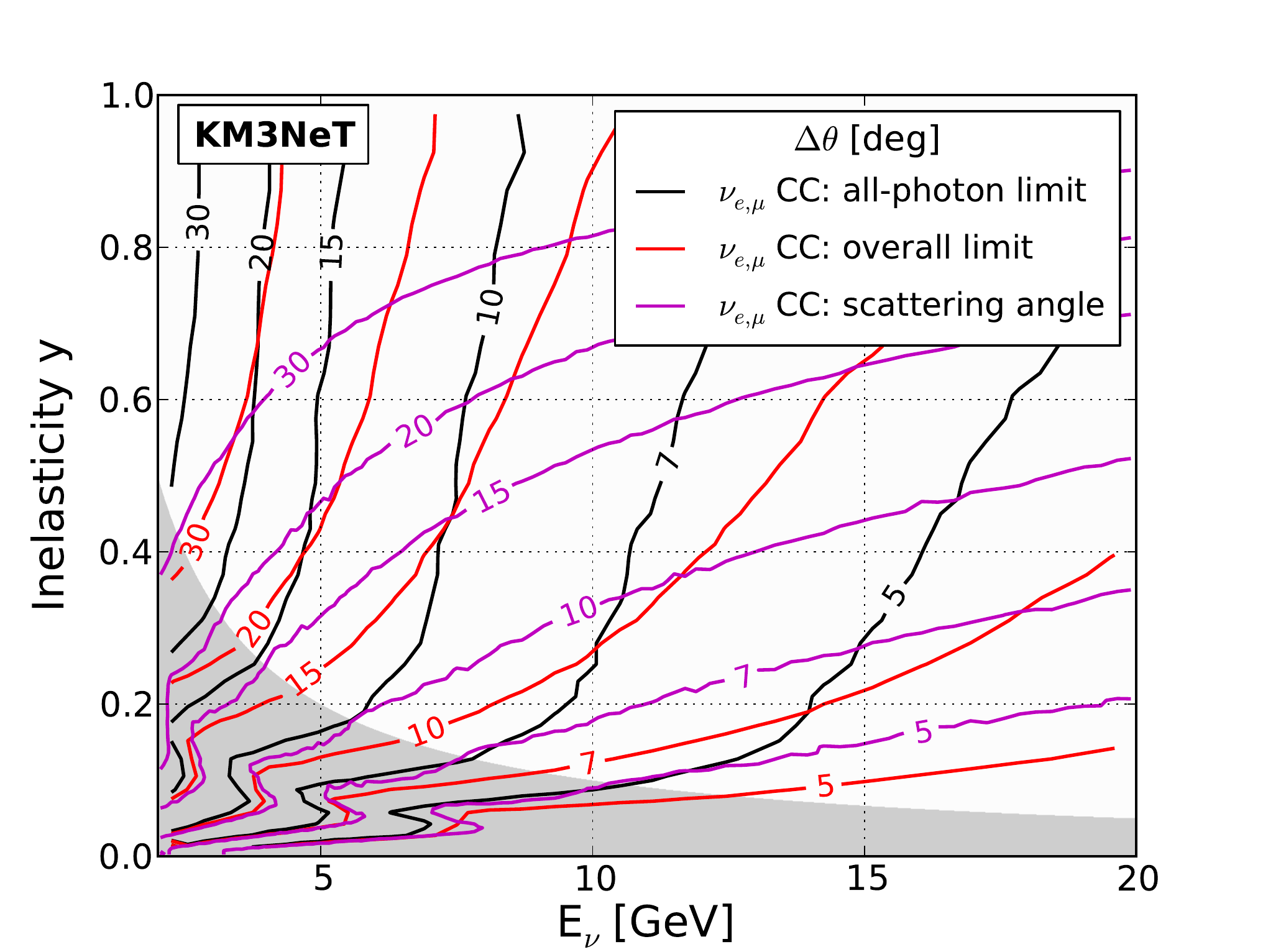}
\end{center}
\caption{Limitations on neutrino direction resolution (68\% quantiles) in $\protect\nuan_{e,\mu}$~CC events as a function of neutrino energy $E_\nu$ and inelasticity $y$, in both the all-photon limit (black) and the overall limit (red).
This is compared to the intrinsic scattering angle between the outgoing lepton and the neutrino (purple). The region $E_h < 1$\,GeV (grey shading) has been calculated using extrapolated hadronic cascade results, and should be interpreted with care. 
The `kinks' in the bottom-left corner are due to a change in recoil mass (different interaction channels) and therefore different values of $E_h/|\vec{p}_h|$.
} 
\label{fig:reso2D_direction_numuANDnue}
\end{figure}

Figures~\ref{fig:reso2D_energy_numuANDnue} and~\ref{fig:reso2D_direction_numuANDnue} show the resulting limits using naive four-momentum conservation on the neutrino energy and direction resolution for $\nuan_e$ and $\nuan_\mu$~CC events as a function of neutrino energy $E_\nu$ and inelasticity $y$. Due to the large differences between resolutions of electrons/muons and hadronic cascades, the neutrino resolutions depend strongly on $y$, and are very similar for $\nuan_e$ and $\nuan_\mu$~CC events. 
The exception is the overall limits on energy resolution, which are shown separately.

\begin{figure*}[tbp]
\begin{center}
\includegraphics[width=0.8\textwidth]{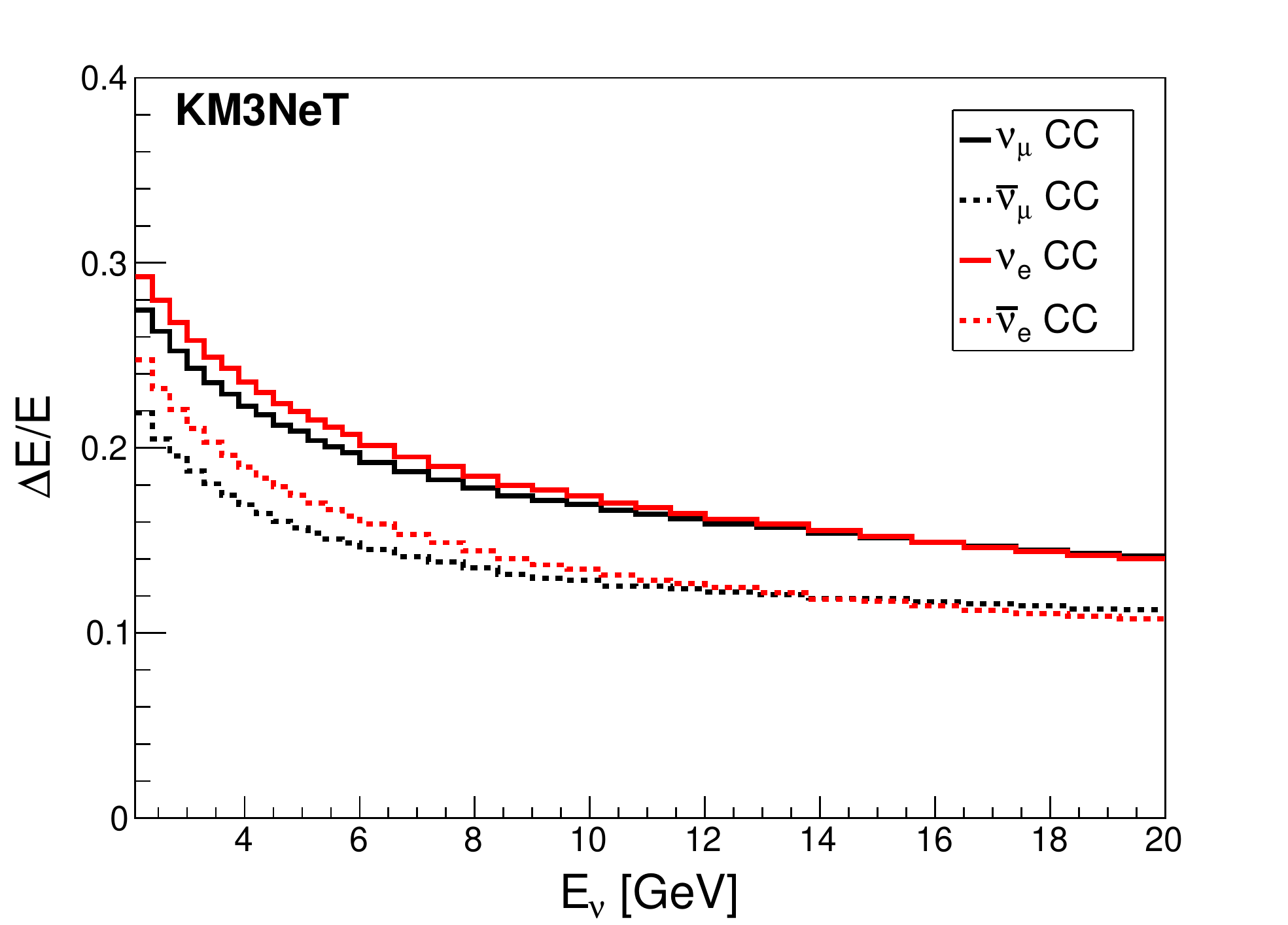}
\end{center}
\caption{Relative energy resolution (RMS) for $\nu_\mu$ (black solid), $\overline{\nu}_\mu$ (black dashed), $\nu_e$ (red solid) and $\overline{\nu}_e$~CC events (red dashed).
} 
\label{fig:energy_reso_numu_nue}
\end{figure*}

\begin{figure*}[tbp]
\begin{center}
\includegraphics[width=0.8\textwidth]{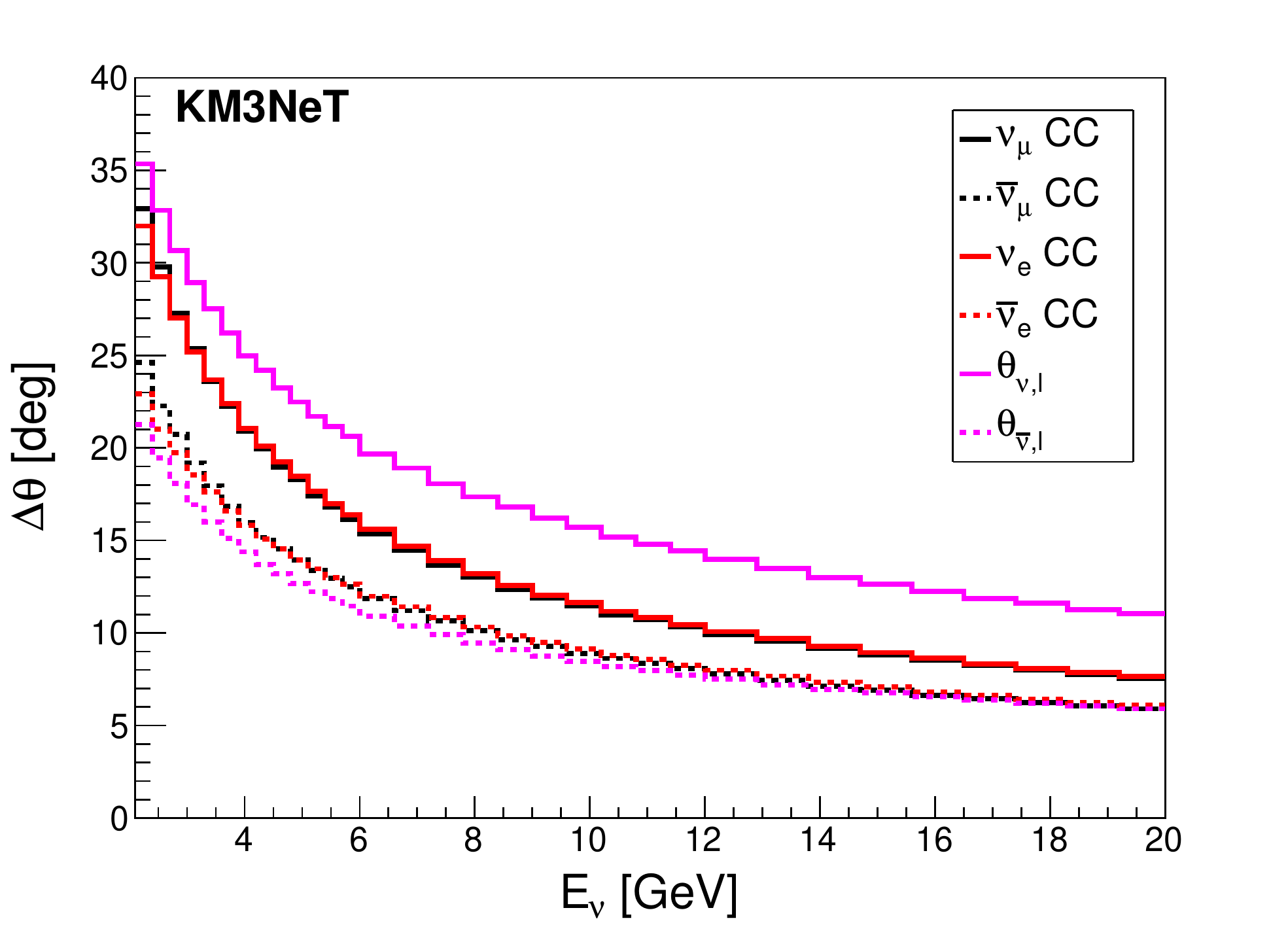}
\end{center}
\caption{Direction resolution (68\% quantiles) for $\nu_\mu$ (black solid), $\overline{\nu}_\mu$ (black dashed), $\nu_e$ (red solid) and $\overline{\nu}_e$~CC events (red dashed). For comparison, the 68\% quantiles of the scattering angle $\phi_{\protect \nuan,\ell}$ between the (anti)neutrino and lepton is shown as purple lines.}
\label{fig:direction_reso_numu_nue}
\end{figure*}

Figures~\ref{fig:energy_reso_numu_nue} and~\ref{fig:direction_reso_numu_nue} show the neutrino energy and direction resolutions for $\nu_e$, $\overline{\nu}_e$, $\nu_\mu$ and $\overline{\nu}_\mu$~CC events as a function of neutrino energy integrated over the corresponding $y$ distributions. Both energy and direction resolutions are better for $\overline{\nu}$~CC than for $\nu$~CC events, since the former have on-average smaller $y$. Because the muon track length is a more reliable energy estimator than the total light yield, the energy resolution for $\nuan_{\mu}$ is slightly better than for $\nuan_e$ (see appendices~\ref{sec:muon_tracks} and~\ref{sec:electron_cascades}). For comparison, in case of measuring only the lepton energy --- and ignoring the hadronic cascade energy --- the resolution is nearly energy independent with $\Delta E/E \approx 0.5$ ($0.3$) for $\nu_{e,\mu}$ ($\overline{\nu}_{e,\mu}$)~CC events. Also, note that while the direction resolution for $\nu$~CC events is significantly better than the scattering angle $\phi_{\nu,\ell}$ between neutrino and lepton, the direction resolution for $\overline{\nu}$~CC events is slightly worse.


\subsection{Using the expected neutrino-lepton scattering angle}
\label{sec:reco_expected_scattering_angle}

Using the hadronic cascade for neutrino direction reconstruction is limited predominantly by the large error $\Delta \theta_h$ on the direction of the hadronic cascade. Two secondary factors are: the resolution of $E_h$, which leads to errors in $\vec{u}_{\nu}$ through eq.~\ref{eq:nu_reco_Ereco_weighing}; and the inability to reconstruct $|\vec{p}_h|$, which necessitates some assumption about $E_h / |\vec{p}_h|$. An improvement therefore might be made by only using $\vec{u}_h$ to define the plane of the interaction, and using the expected median scattering angle med[$\phi_{\nu,\ell}$] between the neutrino and lepton to define the direction in this plane. Such a method will tend to improve the neutrino direction resolution when the relative scatter in the angle $\phi_{\nu,\ell}$ about med[$\phi_{\nu,\ell}$] is small. Here, med[$\phi_{\nu,\ell}$] is calculated from the reconstructed energies $E_{\ell}^{\mathrm{reco}}$ and $E_h^{\mathrm{reco}}$ by setting med[$\phi_{\nu,\ell} ( E_{\ell}^{\mathrm{reco}}, E_h^{\mathrm{reco}})$] to med[$\phi_{\nu,\ell} ( E_{\ell}^{\mathrm{true}}, E_h^{\mathrm{true}})$].

\begin{figure*}[tbp]
\begin{center}
\includegraphics[width=0.8\textwidth]{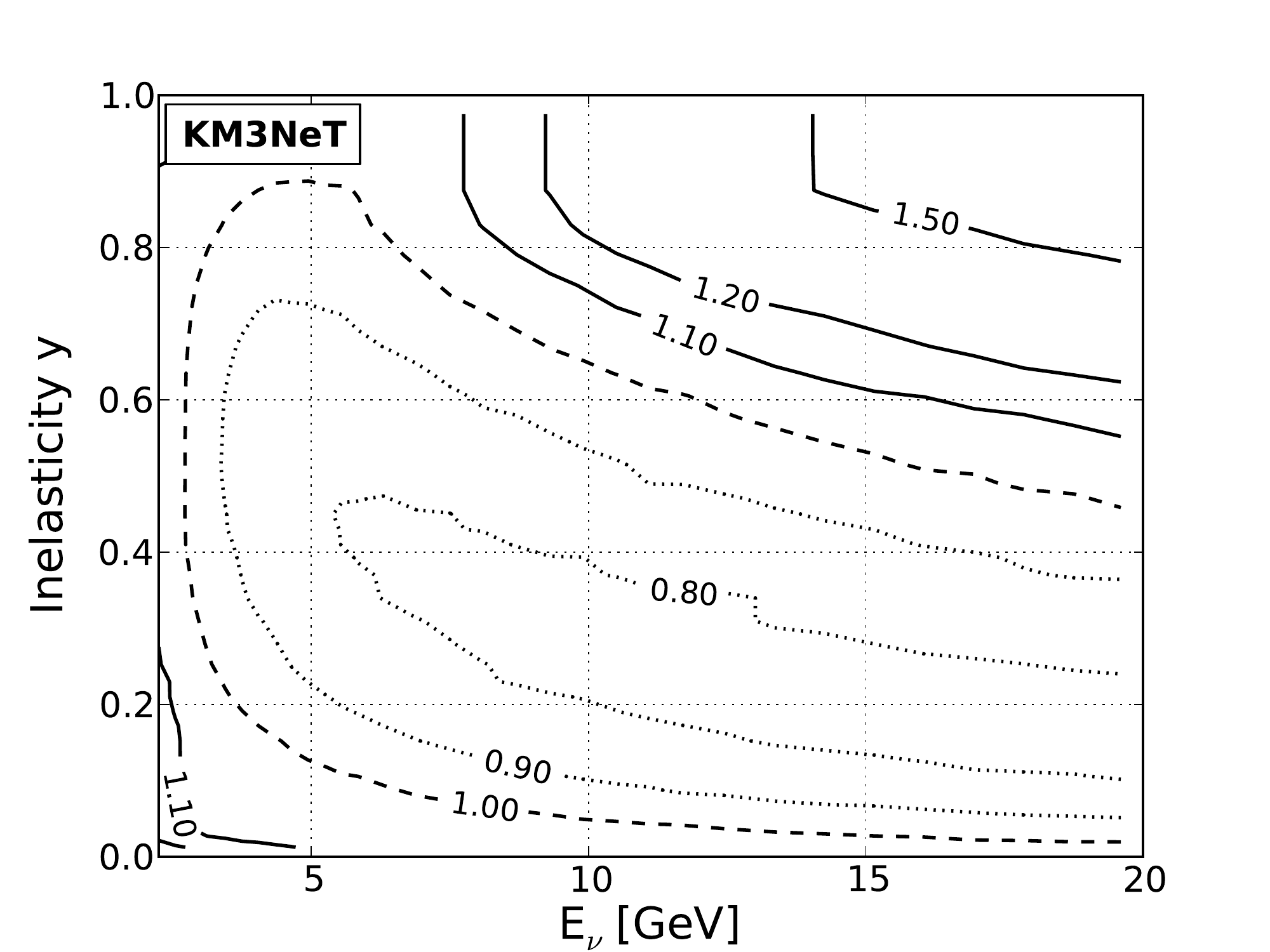}
\end{center}
\caption{Ratio of neutrino direction resolution limits for applying the expected neutrino-lepton scattering angle method over the limits for the naive four-momentum conservation method (cf.\ figure~\ref{fig:reso2D_direction_numuANDnue}) for $\nu_e$~CC events in the overall limit as a function of $E_\nu$ and inelasticity $y$. A ratio $<$ 1 indicates the parameter space where the expected scattering angle method is more accurate.
} 
\label{fig:direction_reso_ratio_kinematic_EPconservation}
\end{figure*}

Figure~\ref{fig:direction_reso_ratio_kinematic_EPconservation} shows the neutrino direction resolution limits for $\nuan_e$~CC events as a function of $E_\nu$ and $y$, relative to those shown in figure~\ref{fig:reso2D_direction_numuANDnue}.\footnote{The resolutions for $\nuan_\mu$~CC are very similar to those for $\nuan_e$~CC events.} For a large region of parameter space in the $E_\nu$--$y$ plane, this method provides a better neutrino direction reconstruction (ratio $< 1$), with up to $\sim 25$\% improvement. It does not perform well at large $y$ due to an intrinsically large scatter in $\phi_{\nu,\ell}$.

\subsection{Treating electron neutrino charged-current events as a single cascade}
\label{sec:resolutions_elecCC_noHitID}

If the detector is not able to distinguish between photons from the electromagnetic and the hadronic cascade in $\nuan_e$~CC events, the reconstruction must treat all photons as coming from a single cascade. This corresponds to applying the method for hadronic cascades of section~\ref{sec:hadronic_cascades} to entire $\nuan_e$~CC events, using the approximate 2.5:1 ratio of atmospheric $\nu_e$ and $\overline{\nu}_e$~CC events in this energy range~\cite{Honda:2015fha,Formaggio:2013kya}.
  
\begin{figure*}[tbp]
\begin{center}
\includegraphics[width=0.8\textwidth]{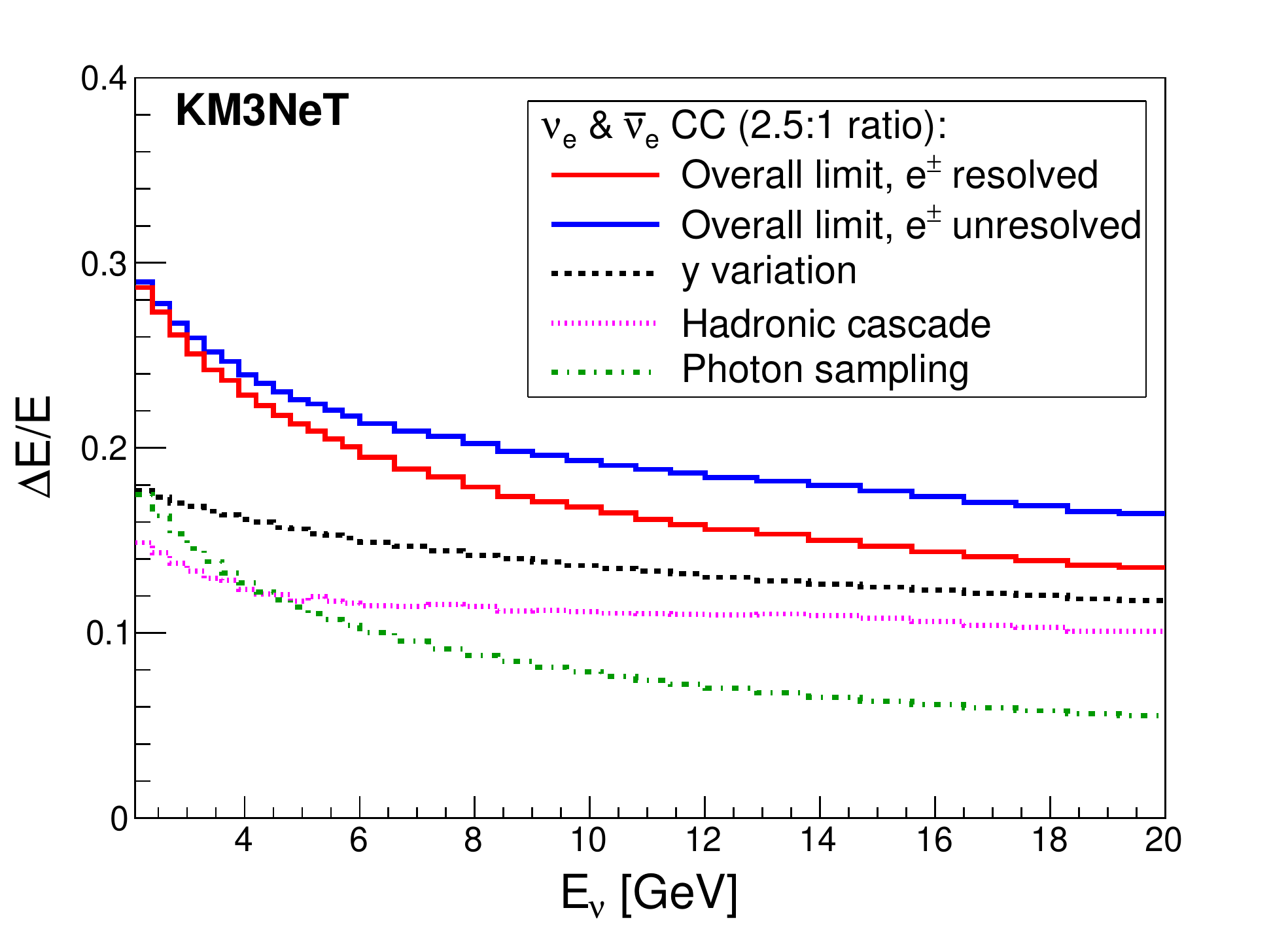}
\end{center}
\caption{Relative neutrino energy resolution (RMS) for $\nu_e$ and $\overline{\nu}_e$~CC events (ratio 2.5:1) in the overall limit, assuming that photons from the outgoing $e^\pm$ and hadronic cascade can (red solid) and cannot (blue solid) be resolved. For the unresolved case, contributions from variation due to unknown inelasticity $y$ (black dashed), hadronic cascade (purple dotted) and photon sampling (green dashed-dotted) are shown separately.
} 
\label{fig:energy_reso_elecCCflux_numu_nue}
\end{figure*}

In figure~\ref{fig:energy_reso_elecCCflux_numu_nue}, the resulting limits on the neutrino energy resolution for $\nu_e$ and $\overline{\nu}_e$~CC events (ratio 2.5:1) are compared to those obtained when the photons from the outgoing $e^\pm$ and hadronic cascade can be resolved. Due to the lack of knowledge about the source of the photons, at $E_\nu=10$\,GeV the energy resolution worsens from $16.8$\% to $19.3$\%. This deterioration emphasises the importance of the capability to identify the electron in $\nuan_e$~CC events in order to allow for more advanced reconstruction procedures.

In the unresolved photon source case, contributions from different sources of variation are separately shown in figure~\ref{fig:energy_reso_elecCCflux_numu_nue}. The variation due to the unknown inelasticity of the interaction is calculated by fixing the number of photons from the electron and hadronic cascades to their mean values and varying $y$ using the true distribution. The variation due to photon sampling is given by the average Poisson error on $N_\gamma$ (eq.~\ref{eq:poisson_var}), summed over both components. The remaining variation is attributed to the contribution from the hadronic state and propagation of the hadronic cascade, and is calculated by subtracting both previously mentioned contributions in quadrature from the overall energy resolution. In addition, when the photon source is unresolved, the energy resolution for a mixed composition of $\nu_e$ and $\overline{\nu}_e$~CC events becomes worse than a simple linear combination of their individual results would suggest, since fluctuations of $N_\gamma$ about the joint mean are necessarily larger. The light yield of $\overline{\nu}_e$ and $\nu_e$~CC events differs by $\sim 10$\% in the considered energy range.

Variations in inelasticity affect the $e^\pm$ unresolved case by smearing the relation between the neutrino energy and the expected $N_\gamma$. However, the effect is still non-zero in the $e^\pm$ resolved case, where fluctuations in $y$ imply that hadronic and electromagnetic energies must be estimated independently. This is why the resolutions for both cases are identical at low energies: when $E_\nu$ is small, almost all light will come from the outgoing $e^{\pm}$; $E_h$ will not be directly observable; and both cases will reduce to measuring $E_e$ only, and relating that to $E_\nu$ based purely on the expected value of $y$.

In the case of direction resolution when the photon source is unresolved, mainly neutrino interactions with small inelasticity ($y \lesssim 0.25$) have a worse resolution. Without the additional knowledge about the source of the photons, the neutrino direction resolution degrades from $11.6^{\circ}$ ($9.2^{\circ}$) to $11.8^{\circ}$ ($10.3^{\circ}$) for $\nu_e$ ($\overline{\nu}_e$)~CC at $E_\nu=10$\,GeV.

\section{Resolution of interaction inelasticity}
\label{sec:inelasticity}

\begin{figure*}[tbp]
\begin{center}
\includegraphics[width=0.8\textwidth]{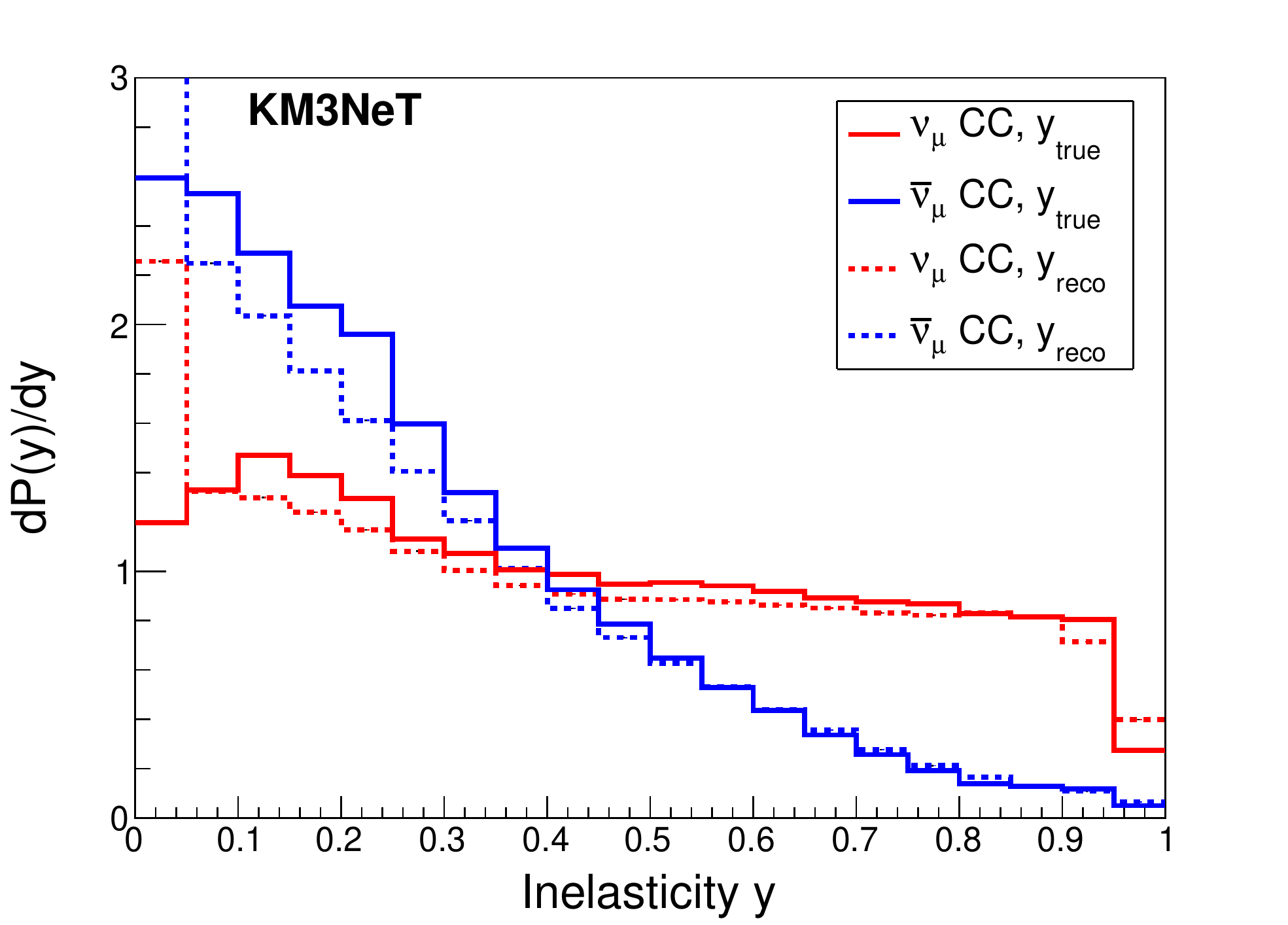}
\end{center}
\caption{
Probability density distributions, $P$, of inelasticity $y^{\rm true}$ (solid) and $y^{\rm reco}$ (dashed) for $\nu_{\mu}$ (red) and $\overline{\nu}_{\mu}$ (blue)~CC events with $E_{\nu}=5$\,GeV. The dip in the last bin (as $y \rightarrow 1$) is caused by the finite muon mass. Due to a fraction of low-energy hadronic cascades producing very few photons, the $y^{\rm reco}$ distribution is shifted to the left, causing the excess in the first bin (as $y^{\rm reco} \rightarrow 0$).}
\label{fig:y_methods}
\end{figure*}

The different differential cross sections of $\nu$ and $\overline{\nu}$ interactions allow them to be statistically separated, which will add significance to a neutrino mass hierarchy measurement~\cite{2013PhRvD..87k3007R}. The energy resolutions obtained in section~\ref{sec:hadronic_cascades} and appendices~\ref{sec:muon_tracks} and~\ref{sec:electron_cascades} can be readily converted into limiting resolutions on the interaction inelasticity $y$ (eq.~\ref{eq:y}), using
\begin{eqnarray}
y^{\mathrm{reco}} & = & \frac{E^{\mathrm{reco}}_h}{E^{\mathrm{reco}}_h + E^{\mathrm{reco}}_{\ell}}. \label{eq:y_reco}
\end{eqnarray}
Figure~\ref{fig:y_methods} shows the $y^{\mathrm{true}}$ distributions for $\nu_{\mu}$ and $\overline{\nu}_{\mu}$~CC events, and the $y^{\mathrm{reco}}$ distributions in the overall limit for $E_{\nu}=5$\,GeV. For $\nuan_e$~CC events, the distributions are very similar. Despite the relatively poor resolution on $E_h$, the differences between the $y^{\mathrm{reco}}$ and $y^{\mathrm{true}}$ distributions are small. Partly, this is due to the lack of structure in the $y$ distribution, particularly for $\nu_{e,\mu}$ interactions, and partly because the error in $y^{\mathrm{reco}}$ tends to zero with both $E_h$ and $E_e$, as can be seen from eq.~\ref{eq:y_reco}.

\begin{figure*}[tbp]
\begin{center}
\includegraphics[width=0.8\textwidth]{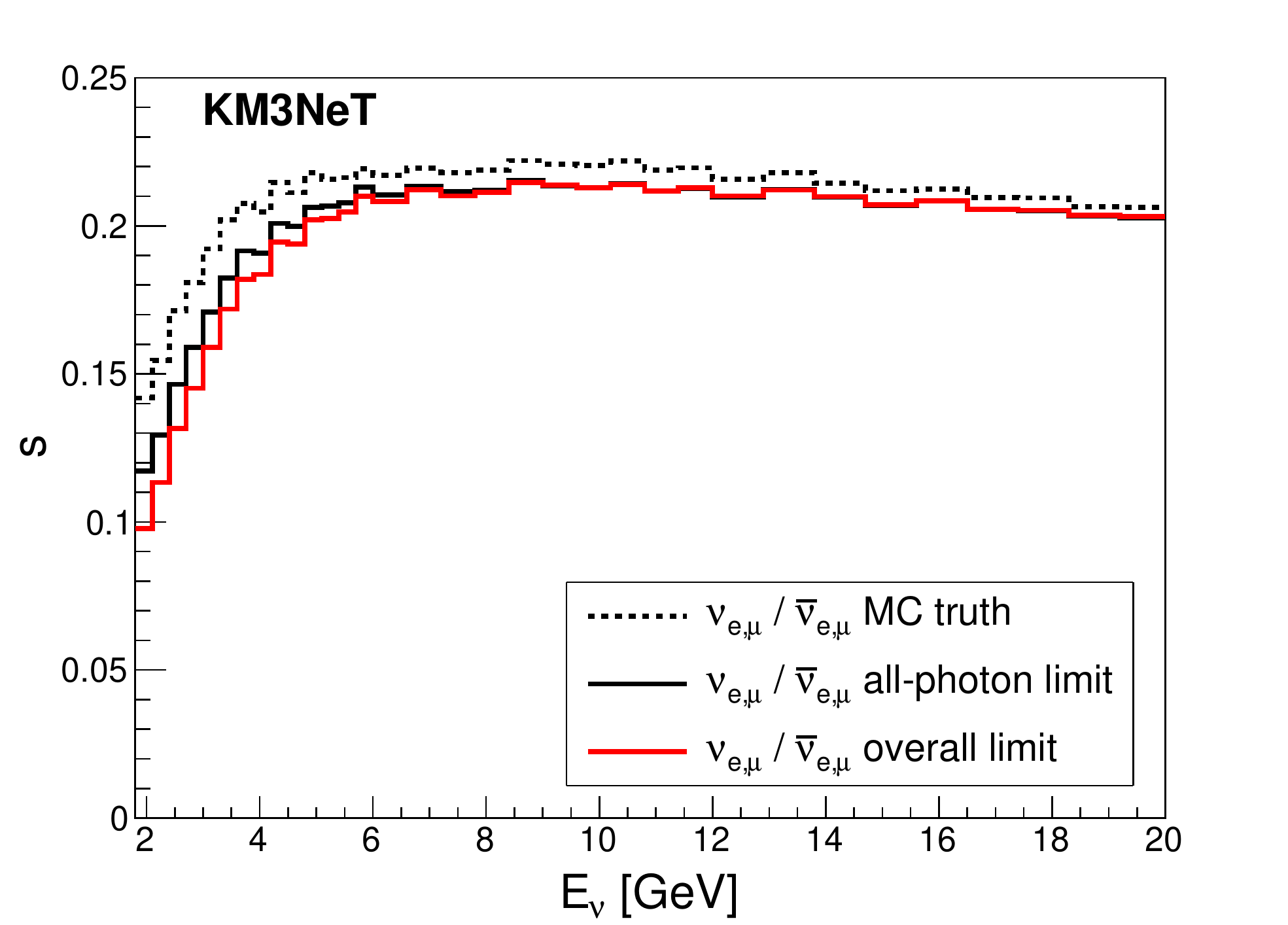}
\end{center}
\caption{Separation power $s$ (eq.~\ref{eq:separation_power}) between $\nu$ and $\overline{\nu}$, using the Monte Carlo truth distributions $P(y^{\mathrm{true}})$, and reconstructed distributions $P(y^{\mathrm{reco}})$ in the all-photon and overall limits. 
}
\label{fig:y_resolution}
\end{figure*}

The statistical resolution between the $y$ distributions from neutrinos $P_{\nu}(y)$ and antineutrinos $P_{\overline{\nu}}(y)$ can be characterised using the correlation coefficient, $c$, by defining the separation power $s$ as
\begin{eqnarray}
s \equiv 1-c & = & 1 - \frac{\int_0^1 P_{\nu}(y) P_{\overline{\nu}}(y) \mathrm{d}y}{ \sqrt{ \int_0^1 P_{\nu}^2(y) \mathrm{d}y \int_0^1  P_{\overline{\nu}}^2(y) \mathrm{d}y }}. \label{eq:separation_power}
\end{eqnarray}
Perfect separation is indicated by $s=1$, while no separation corresponds to $s=0$. The degradation due to limiting accuracies on a detector's ability to distinguish between $\nu$ and $\overline{\nu}$ is given by comparing $s$ calculated using $y^{\mathrm{true}}$ and $y^{\mathrm{reco}}$. Following the same procedure as in section~\ref{sec:reco_energy_momentum_conservation}, figure~\ref{fig:y_resolution} shows the separation power as a function of neutrino energy both with and without reconstruction errors. Above $E_{\nu} = 5$\,GeV, where the intrinsic separation power is highest, the relative reduction in separation power for both all-photon and overall limits is $5$\% or less. 

\section{Discussion}
\label{sec:discussion}

\subsection{Comparison to results using full simulations of KM3NeT/ORCA}
\label{sec:discussion_comparison_ORCA}

The estimated resolutions of the ORCA detector have been published in the Letter of Intent for KM3NeT 2.0~\cite{KM3NeT_LoI}. These results are based on the full simulation chain, which includes the confounding effects of the optical backgrounds, detector layout, PMT response, event triggering and selection, and uses events that are not necessarily fully contained in the detector.

Both $\nuan_\mu$ and $\nuan_e$ CC event reconstruction methods described in ref.~\cite{KM3NeT_LoI} aim to identify the direction of the outgoing lepton, with a limiting resolution for the neutrino direction of the scattering angle $\theta_{\nu,\ell}$ (figure~\ref{fig:direction_reso_numu_nue}). Accounting for different presentation methods (space angle vs.\ zenith angle, median vs.\ 68\% quantile), the resulting direction resolutions (figures 68, 81 and 82 in ref.~\cite{KM3NeT_LoI}) are very close to this limitation. Thus, all prospects for improving the resolution using information from the accompanying hadronic cascade discussed in section~\ref{sec:neutrinos} apply, and the resolution is not downgraded significantly in the case of a detector with a larger spacing between photon sensors. Furthermore, it indicates that the direction resolution of ORCA for $\mu$ and $e$ is not significantly degraded by the aforementioned confounding effects. Indeed, the electron direction resolution of ORCA (figure 82 of ref.~\cite{KM3NeT_LoI}) is close to the `1D' limit (figure~\ref{fig:elec_dir_cos_results} in appendix~\ref{sec:electron_cascades}) presented in this paper.

\begin{figure*}[tbp]
\begin{center}
\includegraphics[width=0.8\textwidth]{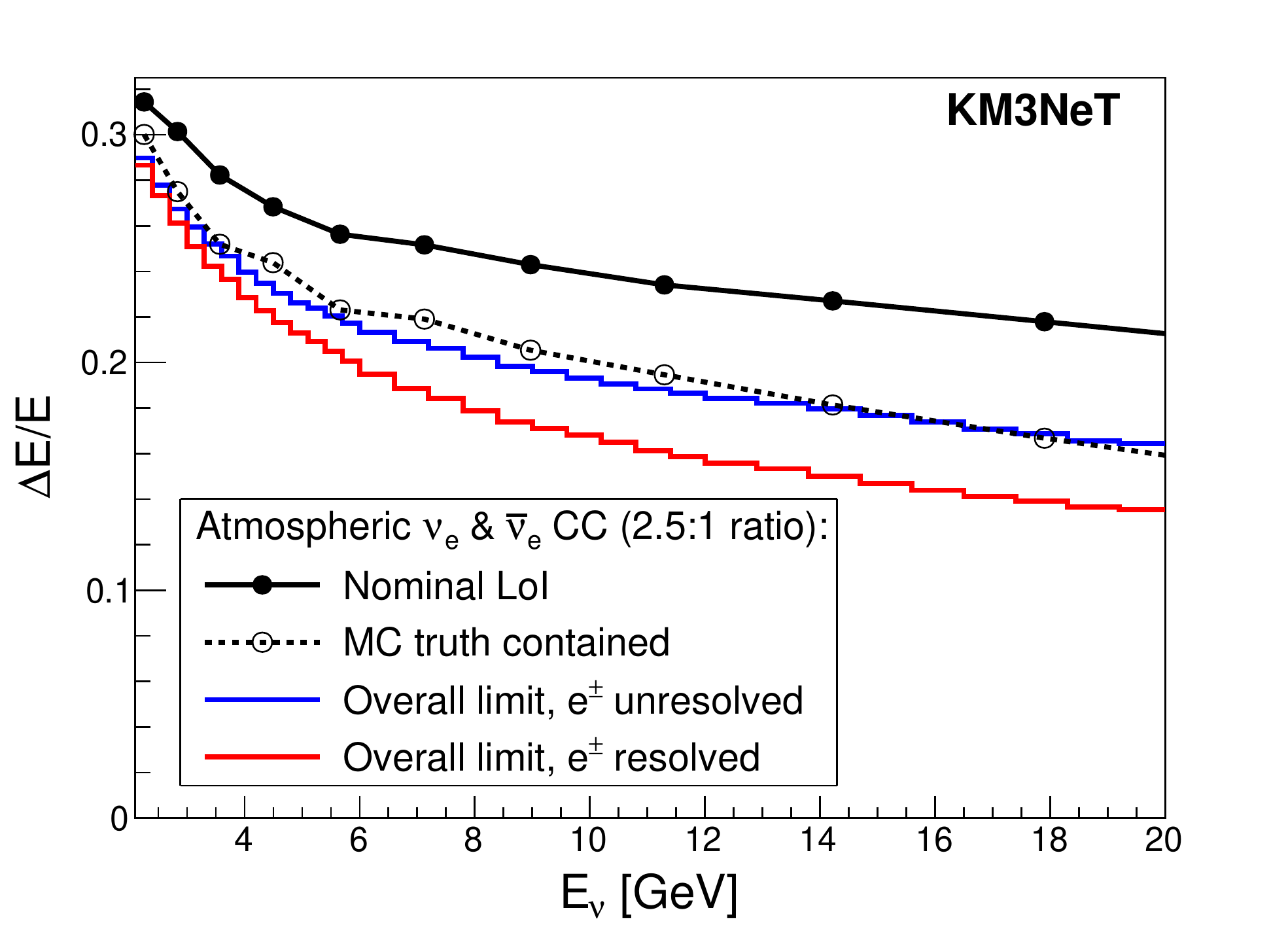}
\end{center}
\caption{
Relative energy resolution (RMS) for KM3NeT/ORCA, calculated using a full detector simulation on atmospheric $\nu_e$ and $\overline{\nu}_e$~CC events.
Results are shown using both the event sample from ref.~\cite{KM3NeT_LoI} (`nominal LoI'), and one based on strict containment criteria on Monte Carlo truth neutrino interaction vertex and direction (`MC truth contained') \cite{JHofestaedt:2016PhD}.
This is compared to the overall limiting resolutions for $\nu_e$ and $\overline{\nu}_e$ CC events (ratio $2.5$:$1$) calculated here, both for the `$e^\pm$ resolved' and `$e^\pm$ unresolved' case, as discussed in section~\ref{sec:resolutions_elecCC_noHitID}.
}
\label{fig:orca_comparison}
\end{figure*}

A comparison to the performance of the energy reconstruction for $\nuan_e$ CC events is shown in figure~\ref{fig:orca_comparison}.
Besides the resolution for the nominal event sample from ref.~\cite{KM3NeT_LoI}, the influence of poorly contained events is eliminated by applying a strict containment criterion based on Monte Carlo truth information. These resolutions are compared to both the $e^\pm$ resolved and $e^\pm$ unresolved case of $\nuan_e$~CC energy reconstruction limits described in section~\ref{sec:neutrinos}. The event selection using strict containment cuts shows an accuracy comparable to the limiting $e^\pm$ unresolved accuracy --- this is not unexpected, since event reconstruction in a real detector cannot perfectly differentiate between photons from the outgoing lepton and from the hadronic cascade.

In the case of $\nuan_\mu$ CC events, the length of the muon track makes it difficult to define a fully contained event for a significant range of the parameter space, and the results are currently too strongly affected by selection effects to make comparisons useful.

\subsection{Applicability to in-ice detectors}
\label{sec:ice_pingu}

The simulations in this paper have been performed in seawater, and assumptions on reconstruction methods are most applicable to the techniques proposed for ORCA. However, since ice and seawater are predominantly composed of H$_2$O, both primary neutrino interactions and subsequent particle propagation are similar. The refractive index of ice is also similar to that of seawater at optical and near-UV wavelengths (between $1.3$ and $1.4$)~\cite{2007APh....27....1R,2006JGRD..11113203A}. Hence, limits due to different hadronic final states and particle propagation (i.e.\ all-photon limits) will therefore be almost directly applicable to in-ice experiments.  Since ice has a lower absorption coefficient than seawater, the photon sampling effects on the energy reconstruction of electrons and hadronic cascades will be significantly lower than those presented here, though perhaps could be calculated with a simple scaling factor. Limits on the neutrino energy resolution (figure~\ref{fig:reso2D_energy_numuANDnue}) however are dominated by fluctuations due to different hadronic states (cf.\ section~\ref{sec:hadron_energy}) and not by limited photon statistics. For direction reconstruction, the photon sampling effect in ice is difficult to estimate: a much higher scattering coefficient reduces the number of unscattered photons, but as discussed above, scattered photons do contain some directional information.

Therefore, it is expected that both all-photon and overall limits on the energy reconstruction accuracy will apply equally well in ice, while direction reconstruction limits are strictly applicable only in the all-photon case.

\subsection{Comparison to estimates in the literature}
\label{sec:literature_comparisons}

The limiting resolutions derived here can be used to assess the validity of several estimates of the sensitivity of PINGU, and hence approximately ORCA, to the neutrino mass hierarchy. Some, such as the calculations of refs.~\cite{2014JHEP...03..028B,Winter2016250,2015PhRvD..91g3011C}, use the published resolutions from the experiments, for which the comparisons above apply. Others, in particular refs.~\cite{2013JHEP...02..082A,2013PhRvD..88a3013W,Ge:2013ffa}, use their own resolution estimates, or approximations of the published results. None of these authors take into account the correlations between energy and direction resolutions, or the differences between $\nu$ and $\overline{\nu}$ interactions.

The neutrino direction resolutions from refs.~\cite{2013JHEP...02..082A,2013PhRvD..88a3013W,Ge:2013ffa} are compatible with the limits presented in this work. However, direction errors are often~\cite{2013JHEP...02..082A,2013PhRvD..88a3013W} assumed to be Gaussian in zenith angle and not in space angle (the limits presented in this work are functions of space angle).

The energy resolutions assumed by these authors~\cite{2013JHEP...02..082A,2013PhRvD..88a3013W} range from 15\% to 35\%, and tend to become incompatible with the limits (cf.\ figure~\ref{fig:energy_reso_elecCCflux_numu_nue}) at low energies, i.e.\ they are too optimistic in this range.

Effects governing neutrino resolution have also been investigated in ref.~\cite{Ge:2013ffa}. However, the authors do not consider the fluctuations in hadronic cascades due to different hadronic final states or particle propagation. Additionally, they overestimate the light yield of hadronic cascades, assuming $f_h = 80$\% (eq.~\ref{eq:LYratio}), independent of energy. This is significantly too large for the considered energies (cf.\ figure~\ref{fig:fit_fh}).
Consequently, they underestimate the contribution due to variation in inelasticity $y$. Furthermore, the neglected effects contribute significantly to the limiting resolutions on neutrino energy. As a consequence, the energy resolutions assumed in ref.~\cite{Ge:2013ffa} are incompatible with the limits derived here.

\subsection{Sensitivity to physical uncertainties}
\label{sec:systematic_uncertainties}

The logic of this paper is to assume that all physical processes are perfectly modelled, and fluctuations within these models are studied. In particular, this applies to the use of {\tt GENIE} for neutrino interaction as well as hadronisation, and {\tt GEANT} for particle propagation. Deviations in this modelling from the truth will have two effects. Firstly, it may lead to systematic shifts between reconstructed and true parameters. These can only act to worse the resolution, but might be accounted for through the use of nuisance parameters in global fits (see e.g.\ section~4.6 in ref.~\cite{KM3NeT_LoI}). Secondly, intrinsic fluctuations about the mean behaviour could either increase or decrease, causing a corresponding worsening or improvement in resolutions. The sensitivity of ORCA to this second effect can be gauged from the relative effects of hadronic state, particle propagation, and photon sampling fluctuations.

As shown in section~\ref{sec:hadronic_cascades}, the energy resolution on the hadronic cascade is dominated by the intrinsic fluctuations in the emitted Cherenkov light due to effects of hadronic state variations. Therefore, experiments such as ORCA might be sensitive to systematic uncertainties in the neutrino interaction and hadronisation process in the range of a few GeV. These processes are very difficult to model. {\tt GENIE} \cite{2010NIMPA.614...87A} uses scaling methods \cite{2009EPJC...63....1Y} to extend {\tt PYTHIA} \cite{2006JHEP...05..026S,2008CoPhC.178..852S} to this region. Nonetheless, there are several remaining discrepancies between experimental data and theory, and work is ongoing to improve the performance of both {\tt GENIE} and other neutrino interaction simulation tools (see e.g.\ ref.\ \cite{2015JPhG...42k5004K}, and references contained therein). The impact of variation in {\tt PYTHIA} parameters has been found to be small for inelasticity measurements \cite{2015JPhG...42k5004K}. As discussed in section~\ref{sec:inelasticity} however, inelasticity is also relatively insensitive to intrinsic fluctuations, and so is not a good indicator of model sensitivity. If experiments such as ORCA are affected by such systematic uncertainties, then data collected with ORCA might also help to constrain the models. 

Determining this sensitivity however would require dedicated studies, which would be beyond the scope of this paper.

\section{Conclusion}
\label{sec:conclusions}

An investigation of the intrinsic limits on resolutions for neutrino events in the 1--20\,GeV energy regime for the KM3NeT/ORCA water Cherenkov detector in the Mediterranean Sea has been performed. Limits on the energy and direction resolution of muon tracks, and of hadronic and electromagnetic cascades have been calculated. These have been combined to derive limits on primary $\nuan_{\mu}$ and $\nuan_{e}$~CC resolutions, and the interaction inelasticity $y$. The results indicate the best reconstruction accuracies achievable with the ORCA detector under the assumption that Cherenkov cones from individual hadronic particles will not be reconstructable. The results are presented for both the benchmark ORCA detector, and in the theoretical case that all photons are detected, which also limits resolutions for a detector with a significantly increased photocathode density.

It is found that the limits imposed by the methods described are close to the ORCA resolutions for $\nuan_e$ CC events obtained using full simulations, indicating that the influence of effects such as natural optical background light and the time- and charge-resolution of KM3NeT PMTs are at most small.

The main result is that the energy resolution of few-GeV neutrinos is primarily dictated by intrinsic fluctuations in the number of emitted Cherenkov photons. The uncertainty due to detecting only a small fraction of them plays only a minor role. Light yield fluctuations are dominated by the fluctuations in the hadronic final state in conjunction with the variation of the interaction inelasticity $y$, emphasising the importance of accurate neutrino interaction models. The neutrino direction resolution is dominated by the large errors in hadronic cascade direction reconstruction, which is mainly driven by detected photon statistics. Due to the significantly worse resolutions for hadronic cascades than for muons and electrons, the neutrino energy and direction errors, are strongly correlated via the inelasticity $y$. The methods allowing an optimum reconstruction which have been identified in this paper thus also depend on $y$. This emphasises the importance of resolving the outgoing lepton in a $\nuan_{e,\mu}$ CC interaction --- a capability which has already been demonstrated for ORCA~\cite{KM3NeT_LoI}. Such a result is also important for the statistical discrimination between $\nu$ and $\overline{\nu}$ using their different $y$ distributions and, hence, increased mass hierarchy sensitivity. Importantly, it has been shown that intrinsic fluctuations do not significantly degrade this discrimination power.

The nature of these limits means that these conclusions will hold for neutrino detection in ice, and for different detector configurations, up to the point that a Super-Kamiokande-style reconstruction of the Cherenkov cone from each individual particle becomes possible. The results of this investigations also explain why the resolutions for ORCA presented in the Letter of Intent for KM3NeT 2.0~\cite{KM3NeT_LoI} do not significantly degrade for larger spacings between optical modules. This motivates an increase in the nominal vertical spacing between optical modules of the ORCA benchmark design from 6\,m to 9\,m. Moreover, the results allow the broader community to make improved estimates of the potential sensitivity of ORCA to the neutrino mass hierarchy, including for alternative detector layouts.
Digitised results are available as supplementary material in the online version of this article.

\acknowledgments

The authors acknowledge the financial support of the funding agencies:
Centre National de la Recherche Scientifique (CNRS), 
Commission Europ\'eenne (FEDER fund and Marie Curie Program),
Institut Universitaire de France (IUF),
IdEx program and UnivEarthS Labex program at Sorbonne Paris Cit\'e (ANR-10-LABX-0023 and ANR-11-IDEX-0005-02), France;
The General Secretariat of Research and Technology (GSRT), Greece;
Istituto Nazionale di Fisica Nucleare (INFN),
Ministero dell'Istruzione, dell'Universit\`a e della Ricerca (MIUR), Italy;
Agence de  l'Oriental and CNRST, Morocco;
Stichting voor Fundamenteel Onderzoek der Materie (FOM), Nederlandse
organisatie voor Wetenschappelijk Onderzoek (NWO), the Netherlands;
National Authority for Scientific Research (ANCS), Romania;
Plan Estatal de Investigaci\'on (refs.\ FPA2015-65150-C3-1-P, -2-P and -3-P, (MINECO/FEDER)), Severo Ochoa Centre of Excellence and MultiDark Consolider (MINECO), and Prometeo and Grisol\'ia programs (Generalitat Valenciana), Spain.

\appendix

\section{Further simulation details}
\label{sec:simulations_appendix}

\subsection{GEANT settings}

{\tt GEANT} was set to track electrons down to their Cherenkov threshold of $0.25$\,MeV, and photons down to $0.4$\,MeV, at which the maximum energy of Compton knock-on electrons is $0.25$\,MeV. The simulation also allows the explicit production and tracking of $\delta$-electrons above $0.25$\,MeV. To ensure the necessary accuracy of electron tracking at low energies, the standard {\tt GEANT 3.21} routine `{\tt gtelec.f}' had to be modified (setting {\tt IABAN} to $0$) to avoid the premature stopping of electrons. Additionally, slow neutrons with kinetic energies below $\sim 100$\,MeV were artificially removed from the simulation to avoid an unphysical number of neutron captures and subsequent nuclei decays. The {\tt GEANT} implementations of both {\tt GHEISHA}~\cite{gheisha} and {\tt FLUKA}~\cite{fluka2005,2014NDS...120..211B} were used for hadronic tracking.

\subsection{G-GHEISHA vs.\ G-FLUKA}
\label{sec:appendix_GHEISHA_vs_FLUKA}

Hadron tracking in {\tt GEANT 3.21} can be performed by either {\tt GHEISHA}~\cite{gheisha} or a preliminary version of {\tt FLUKA}~\cite{fluka2005,2014NDS...120..211B}, often respectively termed `{\tt G-GHEISHA}' and `{\tt G-FLUKA}' to distinguish their implementations in {\tt GEANT} from independent distributions. A comparison of these packages, both with each other, an independent version of {\tt FLUKA}, and experimental data was performed in ref.~\cite{ATLAS_phys_no_086} in the relevant range around $1$--$100$\,GeV. The most relevant observed deviations from expected behaviour were the non-conservation of energy and baryon number by {\tt G-GHEISHA} in pion-nucleon interactions; and an underestimation of up to $30$\% by {\tt G-FLUKA} in the particle multiplicity from pion-carbon interactions below $7.5$\,GeV ({\tt G-GHEISHA} accuracies were of order $10$\%).

In order to characterise the effects of the choice of {\tt G-GHEISHA} or {\tt G-FLUKA}, the resulting distributions of the mean number of detected photons and its standard deviation have been studied. For hadronic cascades with $E_h=10$\,GeV, {\tt G-FLUKA} shows on average a $\sim 6$\% higher light yield (cf.\ figure~\ref{fig:fit_fh}) and $\sim 10$\% less fluctuations than {\tt G-GHEISHA}.

\subsection{Selection of hadronic cascades}
\label{sec:appendix_selecting_had}

\begin{table}[tbp]
\begin{centering}
\begin{tabular}{l|cc}
Component	& Events/energy	& Repetitions/event \\
\hline
Muon tracks		& $1$	& $1000$  \\
Electron cascades	& $1$	& $1000$ \\
Hadronic cascades	& $1000$ & $1000$ {\tt G-FLUKA}, $1000$ {\tt G-GHEISHA}\\
\hline
Energies [GeV]: & \multicolumn{2}{l}{1, 1.5, 2, 2.5, 3, 3.5, 4, 4.5, 5, 6, 7, 8, 9, 10, 12, 14, 16, 18, 20}
\end{tabular}
\end{centering}\caption{Description of the simulations for each component. Each repetition uses a different random number seed, while for hadronic cascades, each event represents a different hadronic final state.} \label{tab:simulations}
\end{table}

The chosen hadronic events were sampled from {\tt GENIE 2.8.4} simulations of $\nuan_{\mu}$~CC events with original neutrino energies between $1$ and $100$ GeV. The first thousand events having a required hadronic energy $E_h$ (defined according to eq.~\ref{eq:ph}) within $1$\% of the nominal values were chosen for each of the energies specified in table~\ref{tab:simulations}. Preliminary studies indicated that for a fixed $E_h$, differences in event properties (mean light yield, particle content) due to a different choice of primary neutrino (flavour, energy) were negligible, in accordance with ref.~\cite{2015PhRvD..92g3014A}.

\begin{figure*}[tbp]
\begin{center}
 \includegraphics[width=0.8\textwidth]{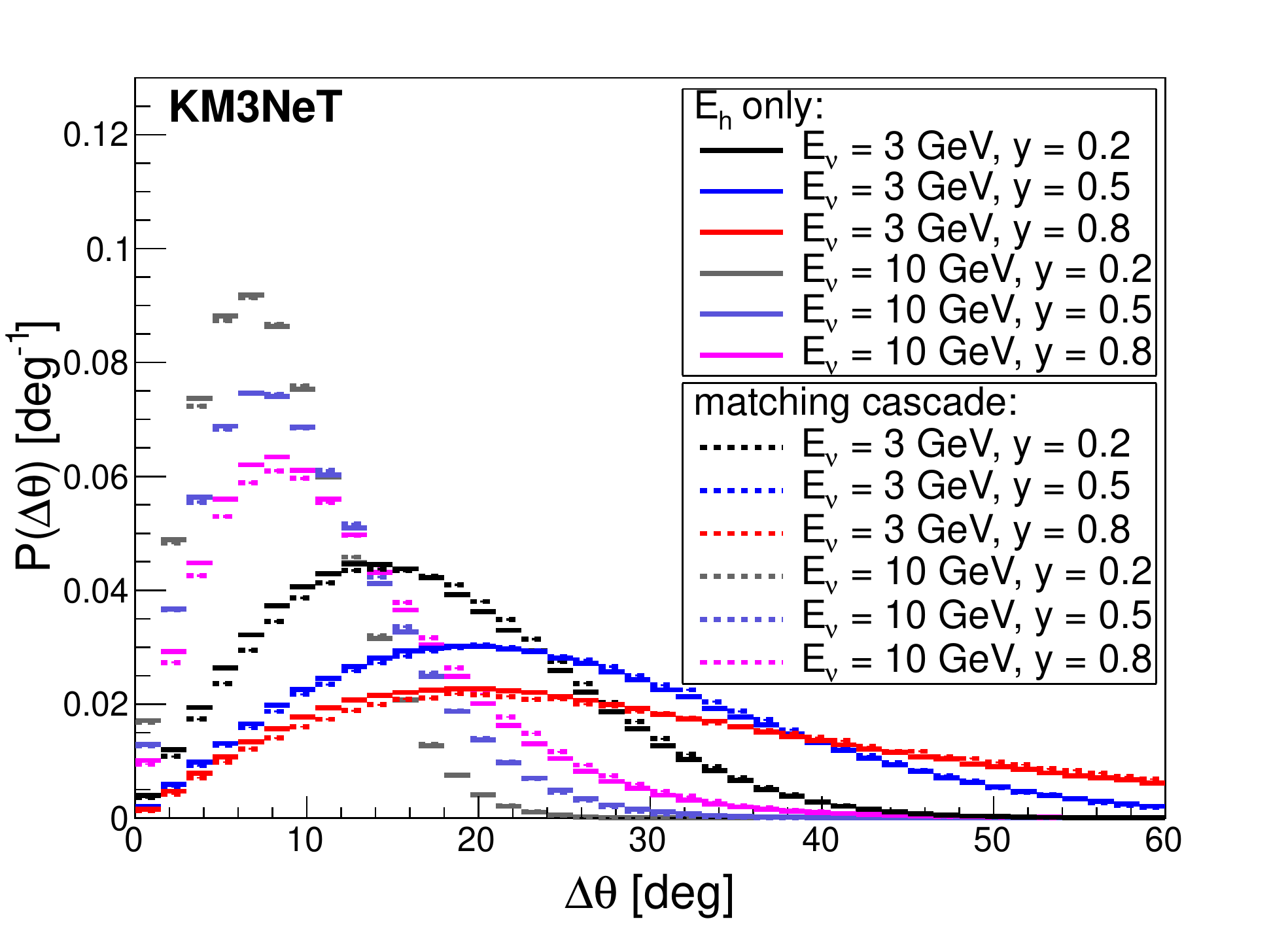}
\end{center}
\caption{
Comparison of neutrino direction resolution $\Delta \theta$ calculated using the methods described in section~\ref{sec:simulations}, when selecting hadronic cascade using $E_h$ only; and when simulating the particular hadronic final state matching each neutrino interaction.}
\label{fig:full_vs_combined}
\end{figure*}

In section~\ref{sec:simulations}, it is noted that more-detailed kinematical properties of neutrino events (four-momentum transfer squared $q^2$ for instance) may be correlated with energy and direction reconstruction errors, and that this association is lost when selecting hadronic final states using $E_h$ only. It was found that the correlation between direction error and $q^2$ is small, with correlation coefficient $|c|<0.05$. The correlation between $N_\gamma$ (cascade energy, section~\ref{sec:hadron_energy}) and $q^2$ is slightly larger, at $c \approx 0.15$. This is still much smaller than the correlation between $N_\gamma$ and the direction error
($c \approx -0.5$), which is accounted for in the results of section~\ref{sec:neutrinos}. Figure~\ref{fig:full_vs_combined} compares the neutrino direction resolution obtained using the method of section~\ref{sec:neutrinos} with that using the exact hadronic final state from each interaction, i.e.\ accounting for any/all correlations, including that with $q^2$. A very good agreement is achieved.

Similarly, investigations have shown that the intrinsic energy and direction resolutions for hadronic cascades induced by $Z^0$ bosons are very similar to that induced by $W^\pm$ bosons \cite{JHofestaedt:2016PhD}, with relative differences in all resolutions below 5\% for $E_h > 3$\,GeV.

A further approximation is required when $E_h$ or $E_{e,\mu}$ do not equal one of the simulated energies of table~\ref{tab:simulations}, in which case the expected properties $\Delta E/E$, $\Delta \theta$ must be interpolated between simulated values. When $E < 1$\,GeV, the values of $\Delta \theta$ and $\Delta E/E$ were set to their values at $1$\,GeV. In the case of energy resolution in the overall limit (including photon sampling), uncertainties due to photon sampling are extrapolated according to the photon statistics ($\sqrt{N_\gamma}$), and uncertainties due to muon track length determination are assumed to be independent of initial muon energy.

The effects of these approximations are included in figure~\ref{fig:full_vs_combined} for $E_{\nu} = 3$\,GeV, $y=0.2$ and $y=0.8$. The agreement with full simulations is only slightly worse.

\section{Muon tracks}
\label{sec:muon_tracks}

\subsection{Energy resolution}
\label{sec:muon_energy_resolution}

In the considered energy range of $1$--$20$\,GeV, muons behave very similarly to minimum ionising particles: they experience a roughly constant energy loss per unit track length, and travel in approximately straight lines. Therefore, their signature in water and ice Cherenkov detectors is a straight track with a nearly uniform luminance. 

This behaviour leads to two obvious methods of muon energy reconstruction: the track length, and the total Cherenkov light yield.

\paragraph{Light yield.}
\label{sec:muon_total_light_yield}
The same principles as for hadronic cascade energy reconstruction discussed in section~\ref{sec:hadron_energy} are used, simplified by the number of emitted photons $N_{\gamma}$ being almost directly proportional to $E_\mu$. Thus, relative all-photon energy reconstruction errors are calculated as
\begin{eqnarray}
\frac{\Delta E_{\mu,N}}{E_\mu} & = & \frac{\Delta N_\gamma}{N_\gamma}. 
\label{eq:muon_tly_perf}
\end{eqnarray}
The sampling errors are given by eq.~\ref{eq:poisson_var}, with $N_{\gamma}^{\rm det} = 22.0 \times (E_\mu - E^{\rm th}_\mu)/\text{GeV}$ being the number of detected photons, and $E^{\rm th}_\mu \approx 157$\,MeV the energy threshold for Cherenkov light emission of muons in seawater. These two sources of variation (all-photon and photon sampling) are independent, and can be added in quadrature to find the overall uncertainty.

\paragraph{Track length.}
\label{sec:muon_energy_range_measurement}
Assuming in all cases that the starting point of a muon track can be resolved perfectly using information from the accompanying hadronic cascade, the general strategy of a muon track length measurement is to detect the endpoint of a muon track using information from unscattered photons, i.e.\ by back-projecting onto the track assuming the Cherenkov angle.

\begin{figure*}[tbp]
\begin{center}
\includegraphics[width=0.8\textwidth]{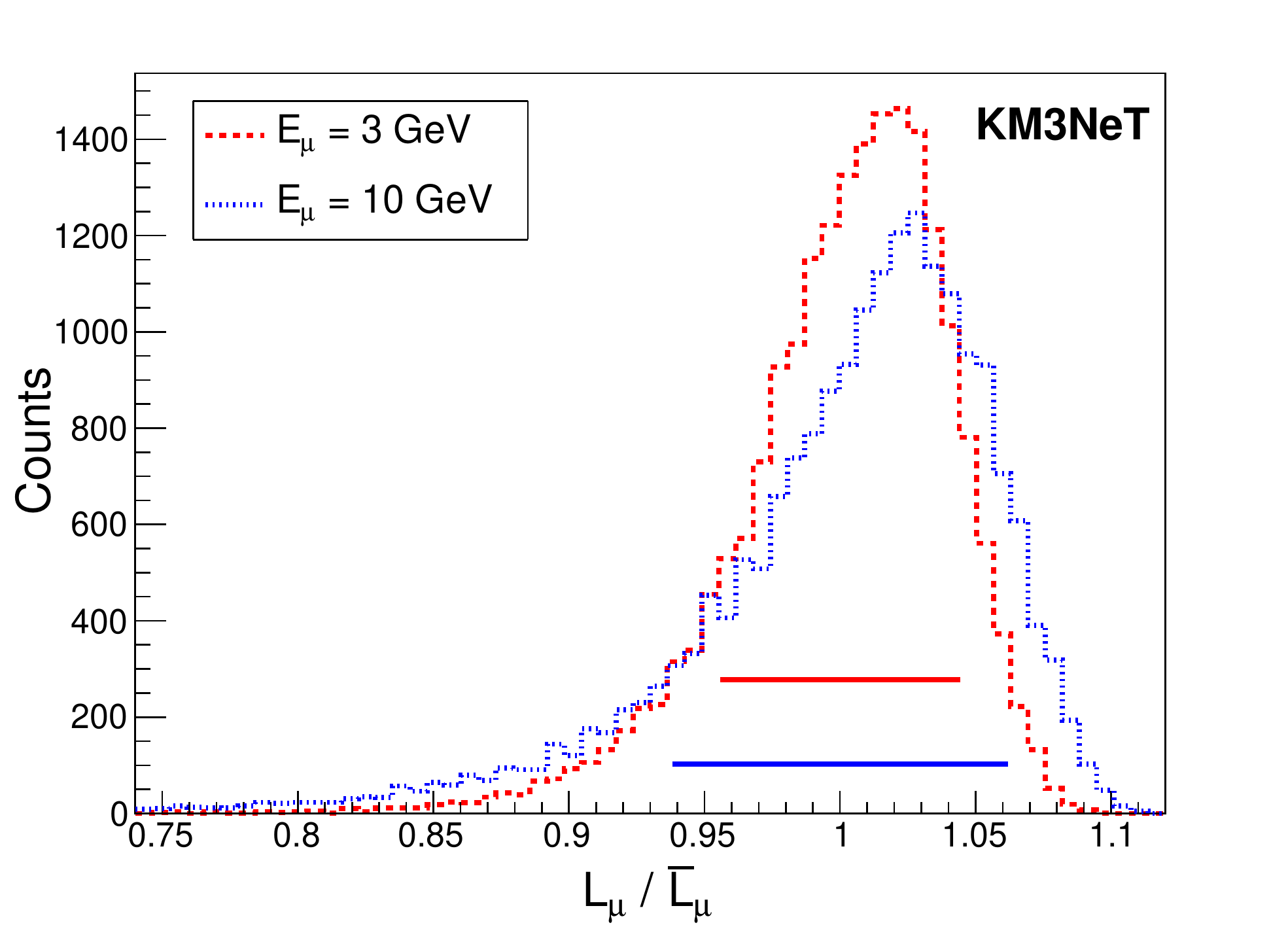}
\end{center}
\caption{Distributions of relative track length, $L_\mu/\overline{L}_\mu$, for muons with initial energies of 3\,GeV (red) and 10\,GeV (blue). Horizontal bars show the mean $\pm$ the RMS.}
\label{fig:method_muon_tracklength}
\end{figure*}

In the all-photon limit, it is assumed that the track endpoint can be determined precisely. Fluctuations in the muon track length $L_{\mu}$ itself will then correspond directly to an error in the energy reconstruction. If $\Delta {L_\mu}(E_\mu)$ is the fluctuation of $L_\mu$, the relative energy resolution in the all-photon limit is given by
\begin{eqnarray}
\frac{\Delta E_{\mu,L}}{E_\mu} & = & \frac{\Delta L_\mu}{L_\mu}. 
\end{eqnarray}
The distribution of the relative track length for $3$\,GeV and $10$\,GeV muons is shown in figure~\ref{fig:method_muon_tracklength}.

In the overall limit, the relatively small number of sampled unscattered photons will limit the resolution on the muon path. The muon track length measurement is then determined by the distance from the starting point to the position where the last detected, unscattered photon is emitted by the muon. The corresponding uncertainty, $\Delta L_\mu^{\rm sampling}$, is estimated from the variation about the mean offset between the position of last emission and the true muon endpoint. For the assumed photon detection probability (figure~\ref{fig:p_det_dir}), this results in roughly $0.1$\,GeV uncertainty on the muon energy. The overall limit on the relative energy resolution is calculated by adding these components in quadrature.

\begin{figure*}[tbp]
\begin{center}
\includegraphics[width=0.8\textwidth]{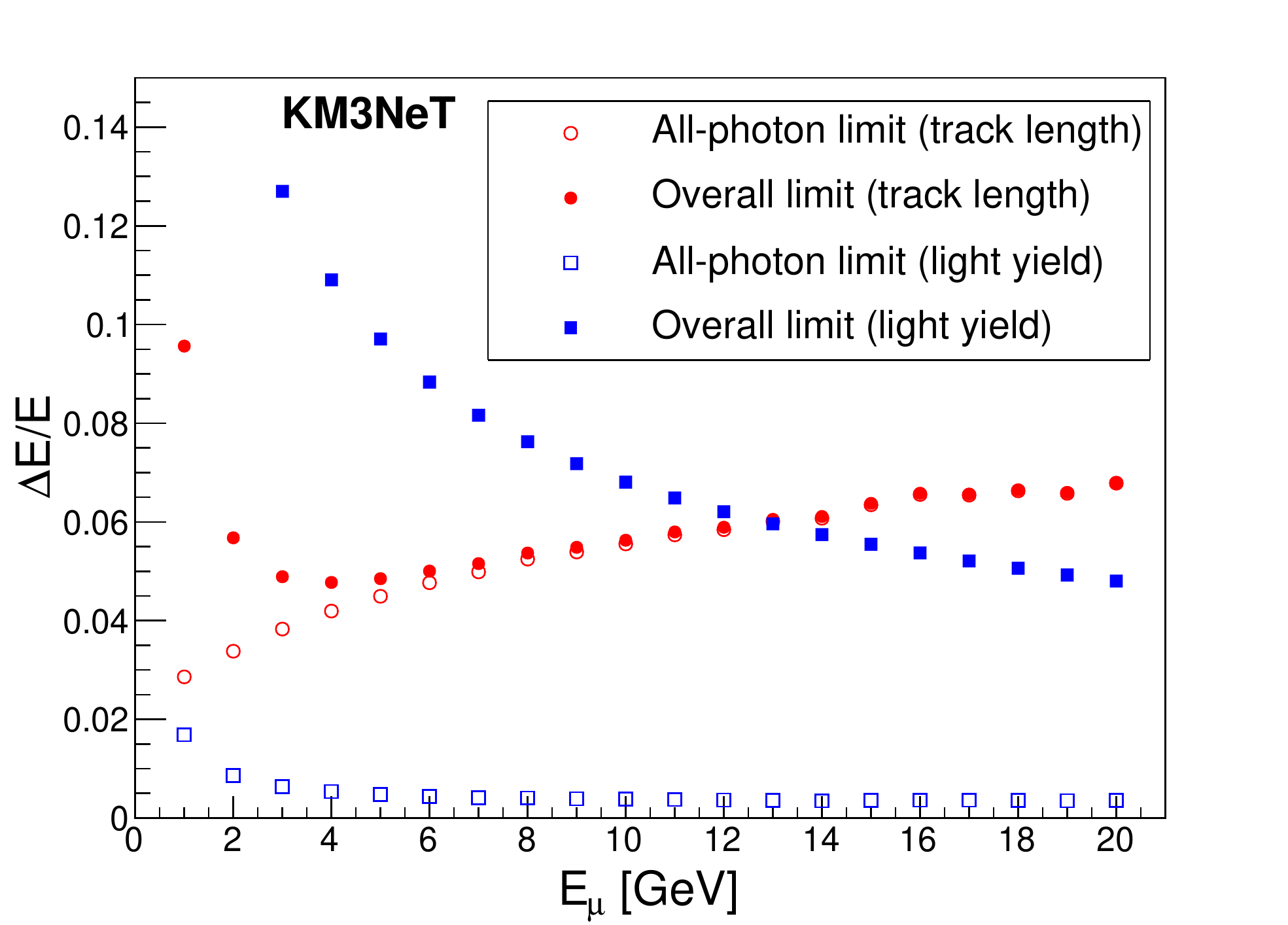}
\end{center}
\caption{
Limits on the relative muon energy resolution (RMS) as a function of muon energy $E_\mu$ for the muon track length method (red circles) and the total light yield method (blue squares). Resolutions are shown for the all-photon limit (hollow markers) and overall limit (filled markers).}
\label{fig:relreso_muon_energy}
\end{figure*}

\paragraph{Results.}
All-photon and overall limits on the relative muon energy resolution for both discussed energy estimates are shown in figure~\ref{fig:relreso_muon_energy}.

In the case of the all-photon limit, fluctuations in total photon emission are almost negligible, whereas those in the muon track length are not, so the total light yield method is more accurate.

In the case of the overall limit, the additional fluctuations due to photon sampling in the total light yield are significantly larger than the uncertainty due to detecting the track endpoint, so that even when combined with intrinsic track length fluctuations, the muon track length method is more accurate for this detector below $13$\,GeV. Therefore, the track length method is used when calculating results for $\nuan_{\mu}$~CC throughout sections~\ref{sec:neutrinos} and~\ref{sec:inelasticity}.

\subsection{Direction resolution}
\label{sec:method_muon_direcion}

In the case of the all-photon limit, under the assumption that the precise emission points of each photon can be reconstructed, in theory it would be possible to measure the initial direction of the muon track immediately after the neutrino interaction, giving perfect direction resolution. Here, a single constraint is introduced to obtain intrinsic limits: that the muon track fit assumes a linear muon path over its entire length, so that the linearity of the track itself limits the accuracy of the direction fit.

The effects of photon sampling in the overall limit to the resolution is unclear, since only a few unscattered photons (with perfect timing and position resolution) need to be detected in order to fully constrain a track fit. It might be possible to obtain such limits by considering the timing uncertainty of the PMTs, or the dispersion characteristics of seawater. However, such limits will be very small indeed, as verified by the very good direction resolution of muon tracks in ANTARES~\cite{2012ApJ...760...53A}. Hence, only the intrinsic fluctuations from the straightness of a muon track itself will be considered, and applied equally to both the all-photon and overall limits.

\begin{figure*}[tbp]
\begin{center}
\includegraphics[width=0.8\textwidth]{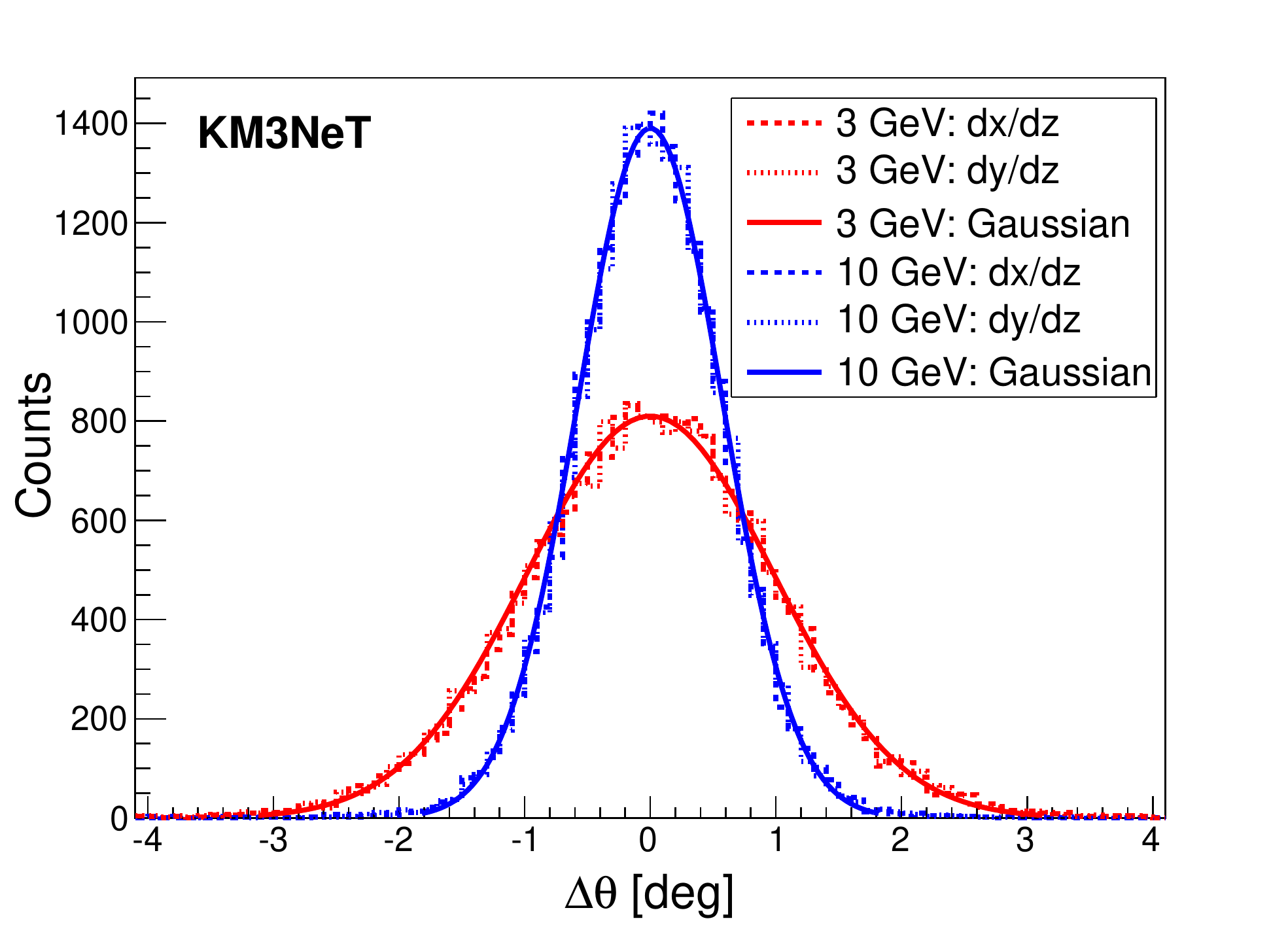}
\end{center}
\caption{Distributions of the slopes d$x$/d$z$ (dashed) and d$y$/d$z$ (dotted) from fits to muons which began traveling along the $z$-axis with an initial energy of 3\,GeV (red) and 10\,GeV (blue). 
Gaussian fits (solid) are also shown.}
\label{fig:method_muon_direction}
\end{figure*}

The slopes d$x$/d$z$ and d$y$/d$z$ from fits to the true muon path (which always begins travelling along the $z$-axis) are used as the errors of such a muon track fit. Their distributions are fitted with Gaussians to derive their standard deviations, $\sigma_{x,y}$. This is shown for $3$\,GeV and $10$\,GeV muons in figure~\ref{fig:method_muon_direction}.

\begin{figure*}[tbp]
\begin{center}
\includegraphics[width=0.8\textwidth]{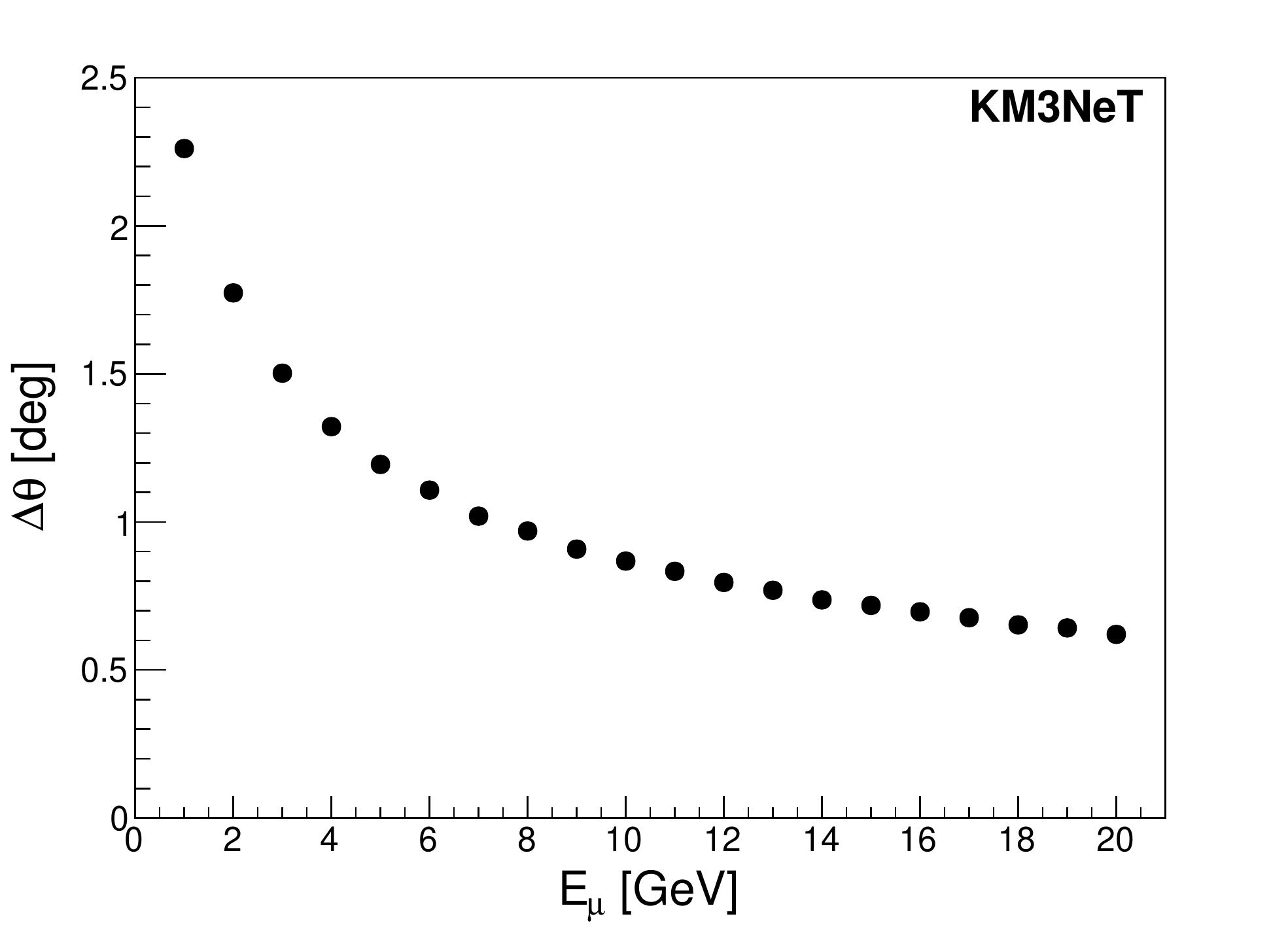}
\end{center}
\caption{
Limiting resolutions on muon direction reconstruction (68\% quantiles) as a function of muon energy $E_\mu$.}
\label{fig:reso_muon_direction}
\end{figure*}

The resulting direction error, $\Delta \theta = \sqrt{2} \sigma_{x,y}$, is both the all-photon and overall limit, and is shown as a function of muon energy in figure~\ref{fig:reso_muon_direction}. Note that these direction errors are dominated by multiple scattering, although the fitting procedure results in a different direction error than more common measures of multiple scattering angles~\cite{pdg}.

\section{Electron cascades}
\label{sec:electron_cascades}

\subsection{Energy resolution}
\label{sec:electron_energy}

\begin{figure*}[tbp]
\begin{center}
\includegraphics[width=0.8\textwidth]{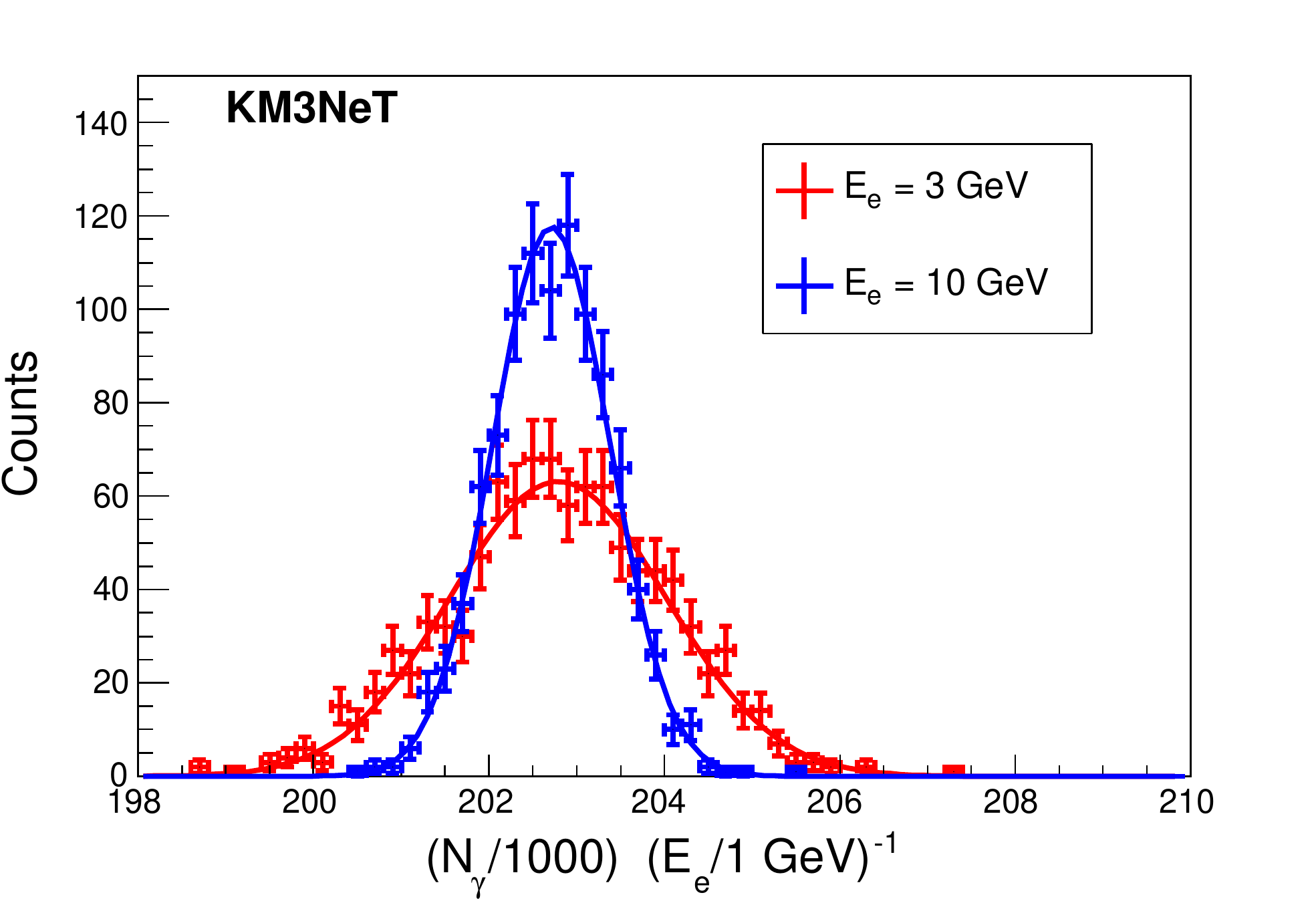}
\end{center}
\caption{
Distributions of the number of emitted photons $N_\gamma$ per GeV of initial energy for $3$\,GeV (red) and $10$\,GeV (blue) electrons (points, with Poisson errors). Gaussian fits are also shown (lines).
} \label{fig:elec_energy_method}
\end{figure*}

The energy of an electron can be best estimated by counting all detected photons produced by its subsequent cascade, which scales almost linearly with the electron energy $E_e$. The resulting fluctuations can be calculated analogously to the total light yield method for muons discussed in appendix~\ref{sec:muon_energy_resolution}. For the all-photon limit, distributions of the total number of detected photons $N_\gamma$ for $3$\,GeV and $10$\,GeV electrons are given in figure~\ref{fig:elec_energy_method}. For the overall limit, Poisson variation is calculated from the total number of detected photons $N_e^{\rm det} = 20.8 \times E_e/\text{GeV}$.

\begin{figure*}[tbp]
\begin{center}
\includegraphics[width=0.8\textwidth]{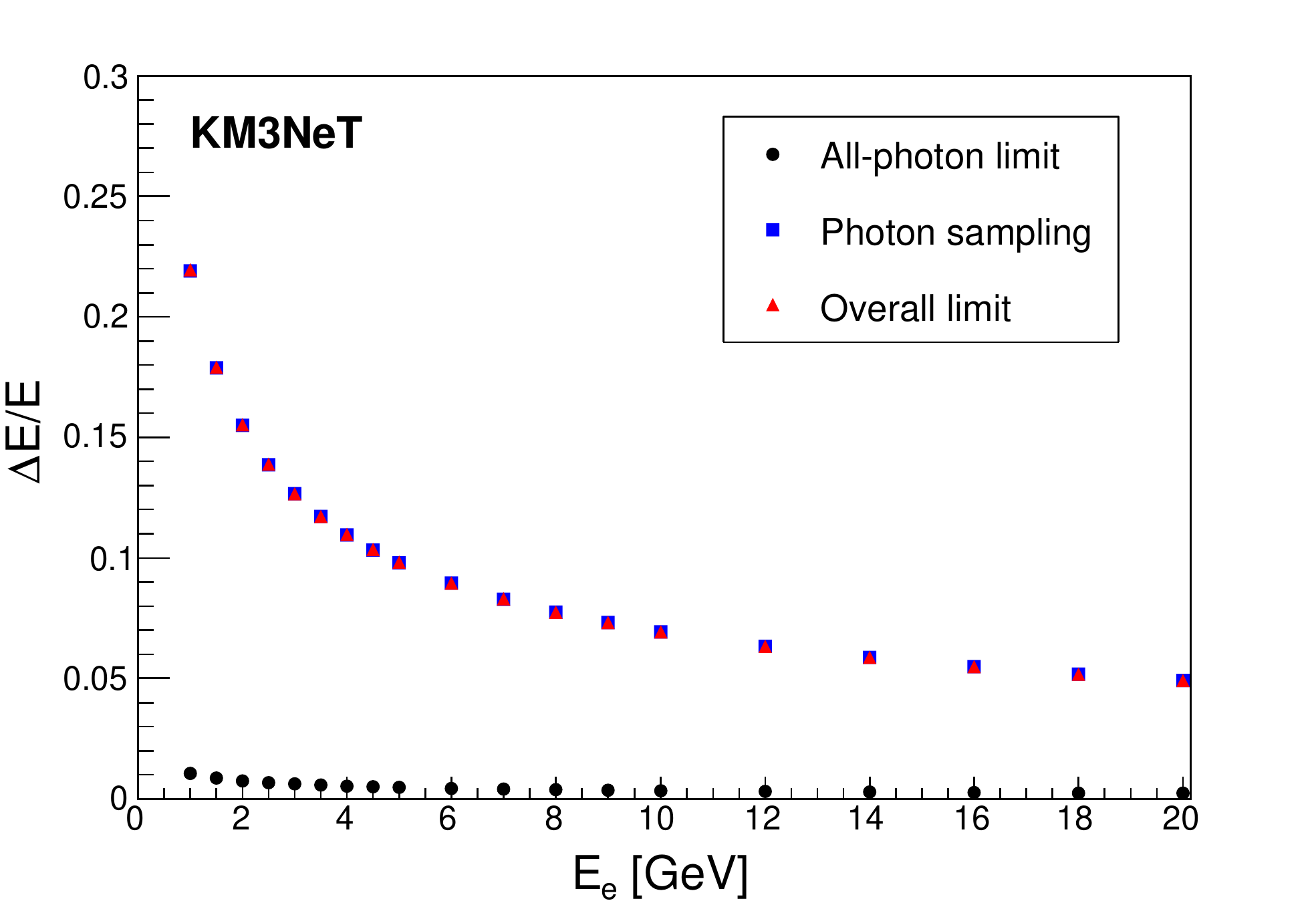}
\end{center}
\caption{Limiting relative energy resolutions (RMS) as a function of electron energy $E_e$, for the all-photon limit (black), the error due to photon sampling (blue, overplotted by red), and for the overall limit (red).} \label{fig:elec_energy_results}
\end{figure*}

The resolutions resulting from these fluctuations are shown in figure~\ref{fig:elec_energy_results}. The dominant uncertainty in the overall limit is due to photon sampling, i.e.\ Poisson fluctuations in the number of detected photons.

\subsection{Direction resolution}
\label{sec:electron_direction}

\begin{figure*}[tbp]
\begin{center}
\includegraphics[width=0.8\textwidth]{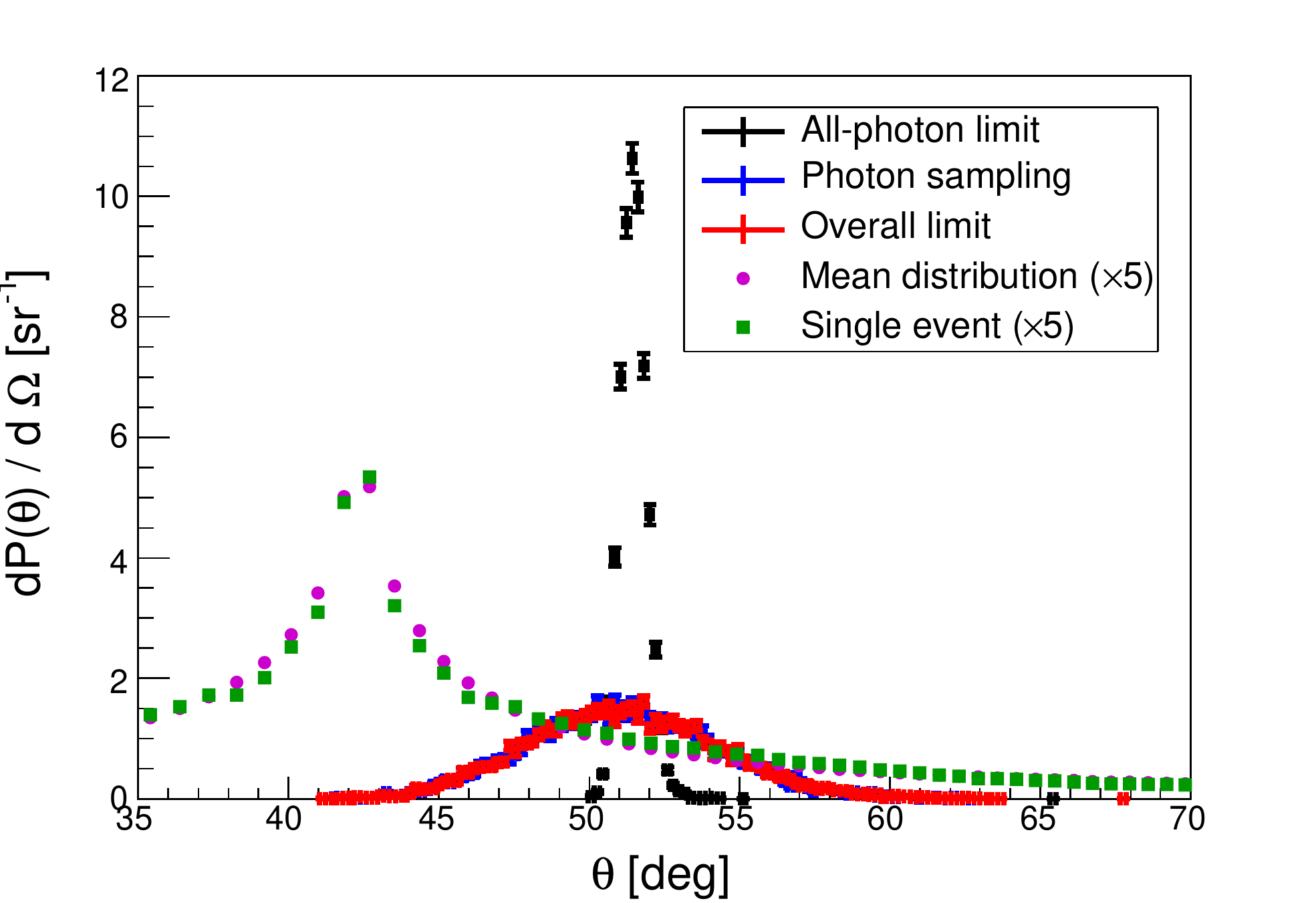}
\end{center}
\caption{
Angular probability distributions of unscattered light from $5$\,GeV electrons, shown for the range of emission angles $\theta$ near the Cherenkov peak at $\theta_C \approx 42^\circ$. The angular distribution averaged over $1000$~cascades (purple circles) is compared to that from a single cascade (green squares); both are multiplied by $5$ for clarity. The means of the angular distributions (calculated between $0^{\circ}$ and $180^{\circ}$) are in the approximate range $50^{\circ}$--$55^{\circ}$, due to there being more photons above the Cherenkov angle than below. The distribution of these means from $1000$~cascades is also shown, calculated using all photons (black), and a sample of photons to give the overall limit (red). The effect due to photon sampling only (blue, overplotted by red) is calculated relative to the mean $\theta$ value for all photons in the cascade.
} \label{fig:elec_angle_cos_method}
\end{figure*}

The method of estimating errors in the reconstructed direction of hadronic cascades presented in section~\ref{sec:hadron_direction} (`2D' method) is motivated by the intrinsic asymmetry of the light pattern about the hadronic cascade axis. As this is not the case for a single electron, an alternative (`1D') method is presented below, which is deemed most appropriate for electron cascades.

Electron cascades tend to produce an azimuthally symmetric distribution of Cherenkov light around the cascade axis, which can be described by a one-dimensional function of the emission angle $\theta$ with respect to the cascade axis. A direction reconstruction can then fit this function to trial cascade axes. Here, the direction error is characterised via fluctuations in the mean values of the $\theta$ distributions. Distributions of $\theta$, and of the mean values, are shown in figure~\ref{fig:elec_angle_cos_method} for 5\,GeV electrons.

\begin{figure*}[tbp]
\begin{center}
\includegraphics[width=0.8\textwidth]{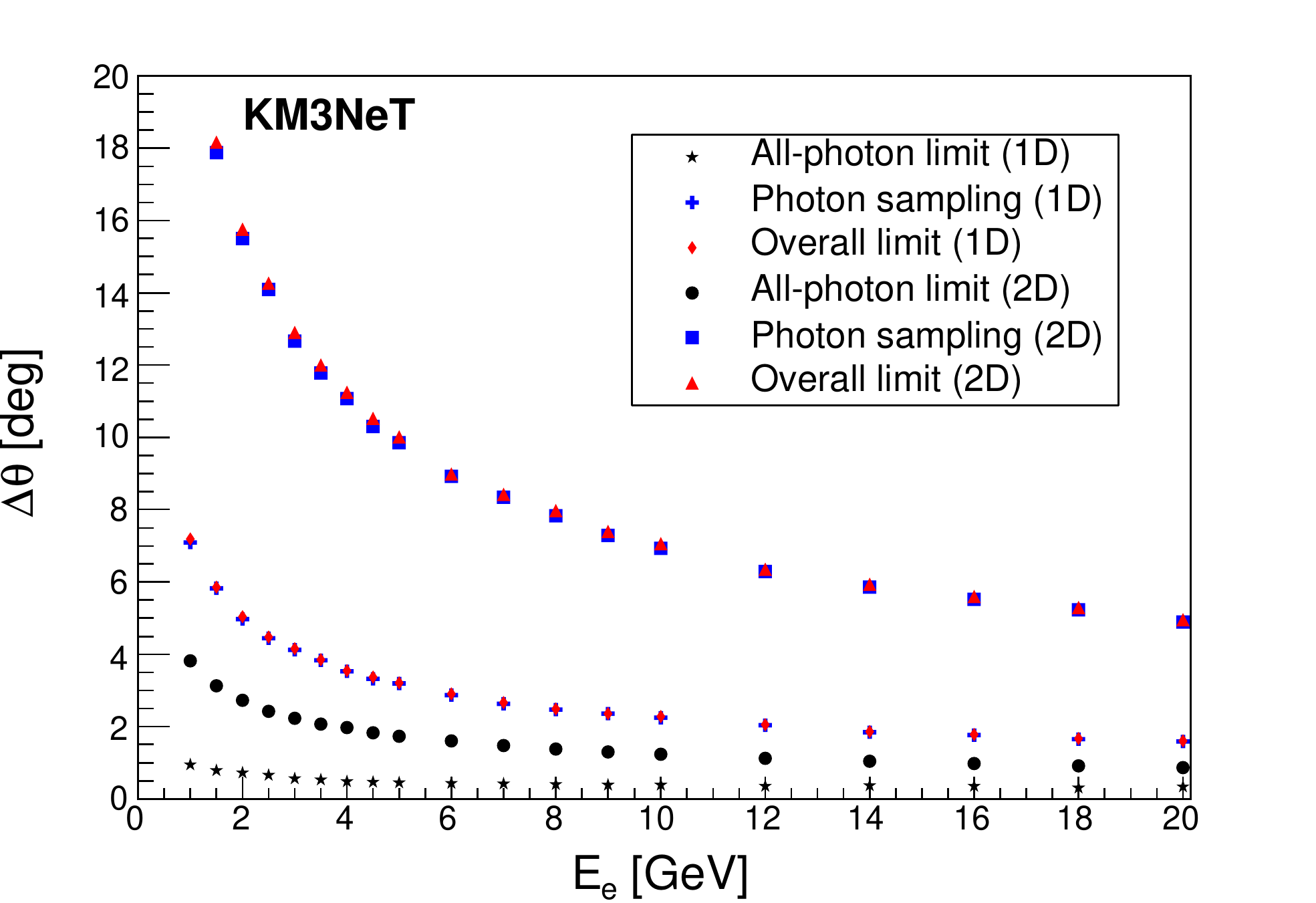}
\end{center}
\caption{
Limiting resolutions on electron direction reconstruction (68\% quantiles), using both the 1D method described in appendix~\ref{sec:electron_direction}, and the 2D method described in section~\ref{sec:hadron_direction}, in the case of the all-photon (black) and overall (red) limits. The effect of photon sampling only is shown in blue (overplotted by red).
} \label{fig:elec_dir_cos_results}
\end{figure*}

Results for the full energy range are shown in figure~\ref{fig:elec_dir_cos_results}. For comparison, the direction error resulting from the 2D method, as used for the hadronic cascade resolution in section~\ref{sec:hadron_direction}, is also shown. Errors due to photon sampling, and hence the overall limits, are dominant over the entire energy range.

The 1D method underestimates fluctuations by using Monte Carlo truth information to define an axis of symmetry, while the 2D method introduces artificial fluctuations in the reconstruction of an azimuthally symmetric photon distribution by allowing a non-uniform distribution in the azimuthal angle to influence the reconstructed direction.

For both the $1$D and $2$D cases, the errors in the all-photon limit are nonetheless small compared to that due to photon sampling, so that the detected photon distribution will be approximately azimuthally symmetric. This is in contrast to the case for hadronic cascades (figure~\ref{fig:hadronic_direction_results}), where the large direction error in the all-photon limit is comparable to that due to photon sampling. For electron cascades therefore, azimuthal symmetry (i.e.\ the $1$D method) could be used to reduce the error due to the small number of sampled unscattered photons ($\sim 10$/GeV), while for hadronic cascades in this energy range it could not be, and the $2$D method must be used. The appropriateness of this choice of methods for both electron and hadronic cascades has been verified by full (and computationally expensive) likelihood reconstruction studies on subsets of data, producing results within $\mathcal{O}(10\%)$ of those shown here.

\phantomsection
\addcontentsline{toc}{section}{References}
\providecommand{\href}[2]{#2}\begingroup\raggedright\endgroup

\end{document}